\documentstyle[twoside,12pt,epsfig,rotating]{article}
\def\greaterthansquiggle{\raise.3ex\hbox{$>$\kern-.75em\lower1ex\hbox{$\sim$}}}
\def\lessthansquiggle{\raise.3ex\hbox{$<$\kern-.75em\lower1ex\hbox{$\sim$}}}
\newcommand{\beq}{\begin{equation}}
\newcommand{\eeq}{\end{equation}}
\newcommand{\beqa}{\begin{eqnarray}}
\newcommand{\eeqa}{\end{eqnarray}}
\newcommand{\ba}{\begin{array}}
\newcommand{\ea}{\end{array}}
\newcommand{\bce}{\begin{center}}
\newcommand{\ece}{\end{center}}

\newcommand{\grts}{\greaterthansquiggle}
\newcommand{\lets}{\lessthansquiggle}

\newcommand{\ra}{\rightarrow}

\newcommand{\fb}{{\rm f\/b}}

\newcommand{\recht} {\begin{picture}(7,7)
                      \linethickness{1.7mm}
                      \put(2,1){\line(0,1){5}}
                      \thinlines
                     \end{picture}
                    } 
\newcommand{\rechtl} {\begin{picture}(7,7)
                      \put(2,1){\framebox(5,5){}}
                     \end{picture}
                    }

\def\noi             {\noindent}

\def\a               {\alpha}
\def\b               {\beta}
\def\t               {\theta}
\def\s               {\sigma}
\def\x               {\chi}
\def\ti              {\tilde}
\def\ee              {$e^+ e^-$}
\def\eeto            {e^+ e^- \to}
\def\sq              {\ti q}

\def\st              {\ti t}

\def\stst            {\st_1\,\bar{\st}_1}

\def\sb              {\ti b}

\def\sbsb            {\sb_1\,\bar{\sb}_1}

\def\stau            {\ti \tau}

\def\staustau        {\stau_1\,\bar{\stau}_1}

\def\sf              {\ti f}

\def\snu             {\ti \nu}
\def\slep            {\ti\ell}
\def\ch              {\ti \x^\pm}
\def\chp             {\ti \x^+}
\def\chm             {\ti \x^-}
\def\nt              {\ti \x^0}
\def\sg              {\ti g}

\def\msg               {m_{\ti g}}
\newcommand{\mst}[1]   {m_{\ti t_{#1}}}
\newcommand{\msb}[1]   {m_{\ti b_{#1}}}
\newcommand{\mstau}[1] {m_{\ti \tau_{#1}}}
\newcommand{\msf}[1]   {m_{\ti f_{#1}}}
\newcommand{\mch}[1]   {m_{\ti \x^\pm_{#1}}}
\newcommand{\mnt}[1]   {m_{\ti \x^0_{#1}}}

\def\tW              {\t_W}
\def\sth             {\sin\t}
\def\cth             {\cos\t}

\def\ptmiss          {p\llap/_{T}}

\def\onehf           {{\textstyle \frac{1}{2}}} 
\def\oneth           {{\textstyle \frac{1}{3}}}    
\def\twoth           {{\textstyle \frac{2}{3}}}
\def\onesq           {{\textstyle \frac{1}{\sqrt{2}}}}

\newcommand{\eq}[1]  {\mbox{(\ref{eq:#1})}}

\newcommand{\etal }   {{\em et~al.}}

\newcommand{\chichi}{\mbox{$\nt_1 c \nt_1\bar{c}$}}
\newcommand{\chipchip}{\mbox{$\chp_1 b \chm_1\bar{b}$}}

\newcommand{\chiz}  {\mbox{$\nt_1$}}
\newcommand{\chip}  {\mbox{$\chp_1$}}
\newcommand{\chim}  {\mbox{$\chm_1$}}

\newcommand{\cc}      {\mbox{$c \overline{c}$}}
\newcommand{\gev}      [1]{\mbox{$\mathrm{  #1 \ GeV      }$}}
\newcommand{\bb}      {\mbox{$b \overline{b}$}}
\newcommand{\abs}[1]{\mbox{$ |#1|        $}}      
\newcommand{\tautau}  {\mbox{$\rm \tau^+ \tau^- $}}
\newcommand{\mm}      {\mbox{$\rm \mu^+ \mu^- $}}
\newcommand{\qq}      {\mbox{$q \overline{q}$}}
\newcommand{\ZZ}        {\mbox{$Z^0 Z^0$}}
\newcommand{\WW}        {\mbox{$W^+W^-$}}
\newcommand{\lumifb}   [1]{\mbox{$\cal L = \invfb{#1}$}}
\newcommand{\invfb}    [1]{\mbox{$\mathrm{  #1 \  fb^{-1}}$}}
\newcommand{\sqrts} {\mbox{$\sqrt{s}$}}

\normalsize
\begin{document}
\bibliographystyle{plain}

\begin{titlepage}

{\topskip -1cm 

\begin{flushright}
UWThPh-1996-66\\
HEPHY-PUB 663/97\\
DESY 97-003\\
hep-ph/9701336\\
\end{flushright}
\vspace*{1.2cm}

\begin{center}

\begin{LARGE}
  Search of Stop, Sbottom, $\tau$-Sneutrino, and Stau \\ 
  at an \\
  $e^+e^-$ Linear Collider with $\sqrt{s}=0.5 - 2$~TeV \\
\end{LARGE}

\vspace{1.8cm}

\begin{large}
  A. Bartl\footnote{bartl@Pap.UniVie.ac.at, 
    $\dagger$ helmut@qhepu1.oeaw.ac.at, 
    $\ddagger$ kraml@hephy.oeaw.ac.at,\\ 
    $\flat$ majer@qhepu1.oeaw.ac.at, 
    $\natural$ porod@Pap.UniVie.ac.at, 
    $\sharp$ andre.sopczak@cern.ch; \\
    http://wwwhephy.oeaw.ac.at/p3w/theory/susy/}$^{\small 1}$, 
  H. Eberl$^{\small \dagger 2}$, 
  S. Kraml$^{\small \ddagger 2}$,\\[1mm]
  W. Majerotto$^{\small \flat 2}$, 
  W. Porod$^{\small \natural 1}$,
  A. Sopczak$^{\small \sharp 3}$ \\[8mm]
\end{large}

{\em (1) Institut f\"ur Theoretische Physik, Universit\"at Wien,
           A-1090 Vienna, Austria } \\[1mm]
{\em (2) Institut f\"ur Hochenergiephysik, \"Osterreichische Akademie der 
           Wissenschaften, A-1050 Vienna, Austria } \\[1mm]
{\em (3) DESY-Zeuthen, D-15738 Zeuthen, Germany} \\ 


\end{center}

\vspace{12mm}
\begin{abstract}

We discuss pair production and decays of stops, sbottoms, 
$\tau$-sneutrinos, and staus in $e^+e^-$ annihilation in the 
energy range $\sqrt{s} = 500$~GeV to $2$~TeV. 
Numerical predictions within the Minimal Supersymmetric Standard 
Model (MSSM) for cross sections and decay rates are presented.
We study the stop discovery potential for $\sqrt{s} = 500$~GeV and 
$10~\fb^{-1}$ integrated luminosity for polarized $e^-$ beams. 
Moreover, we give an estimate of the error of the soft--breaking stop
and sbottom parameters that can be obtained by cross section 
measurements with polarized $e^{-}$ beams.
\end{abstract}

}
\end{titlepage}

\setcounter{page}{2}

\section{Introduction}

If nature is supersymmetric at the weak interaction scale, the
masses of the supersymmetric (SUSY) particles are expected
to be lower than approximately $1$~TeV
\cite{Haber,susy}. The weakly interacting SUSY particles are then 
within the reach of an $e^+e^-$ linear collider with a
center--of--mass energy between $500$~GeV and $2$~TeV. An
 $e^+e^-$ linear collider in this energy range will not only be
a discovery machine for SUSY particles \cite{ee500,eetev,nlc}, but 
will also allow detailed measurements of the underlying 
SUSY parameters \cite{tsuka,feng,finel,daniel}.
The experimental search of SUSY particles
which are relatively light will be particularly important. 
The lighter scalar top quark $\st_{1}$,
the SUSY partner of the top quark, and for
$\tan \beta \,\grts\, 10$ also the sbottom $\sb_{1}$ or the stau 
$\stau_{1}$ may even be the lightest visible SUSY particle 
\cite{Ellis,Altarelli,dreeno,Bartl94}.
The reason is that the Yukawa interactions reduce the soft SUSY
breaking masses of the left and right sfermions, $\sf_{L}$ and $\sf_{R}$,  
of the $3^{rd}$
generation, compared to those of the $1^{st}$ and $2^{nd}$
generation \cite{boer,dreema}, and also induce a mixing which may make one 
mass--eigenstate rather light.
The production cross sections and the decay
rates, and thus the discovery reach of these sfermions show a
distinct dependence on the $\sf_{L}$--$\sf_{R}$ mixing 
angles \cite{eetev,Bartl96}.
If the gluino is heavier than $\st_{1}$ and $\sb_{1}$ the most important 
decay modes of these squarks are those into quarks and 
neutralinos or charginos.

We present results for the production of 
stops, sbottoms, $\tau$-sneutrinos, and staus in
$e^+e^-$ annihilation at energies between $\sqrt{s} = 500$~GeV
and $2$~TeV. We also discuss in detail the decays of these
particles. Furthermore, we present an example for signal selection and 
background rejection for stop
production at $\sqrt{s} =$ 500 GeV and \mbox{$\cal{L} = \invfb{10}$}.
In addition, we investigate the possibility of determining
the masses and mixing angle of stops. If SUSY particles are
experimentally discovered, the measurement of their properties
will be the most important step further. Polarization
of the $e^-$ beam plays an important r\^{o}le. 
As we will show, from measurements of the production cross sections 
with polarized $e^-$ beams we can determine the masses and mixing 
angles of the stops and sbottoms and in turn the underlying soft 
SUSY breaking parameters with good precision. 
Knowing the latter we will be able to test the theoretical hypotheses 
about the SUSY breaking mechanism.

We perform our calculations in the framework of the 
Minimal Supersymmetric Standard Model (MSSM) \cite{Haber,susy}. 
This contains the Standard Model (SM)
particles, sleptons $\slep^\pm$, sneutrinos $\snu_\ell$,
squarks $\sq$, gluinos $\sg$, two pairs of charginos $\ch_i$, 
$i = 1, 2$, four neutralinos $\nt_k$,
$k = 1,\ldots,4$, and five Higgs particles $h^0$, $H^0$, $A^0$,
$H^\pm$ \cite{Haber,susy,Gunion}.
The phenomenology of stops, sbottoms, staus, and $\tau$-sneutrinos,  
and their decay products is determined by the following parameters: \\
\begin{description}
\vspace{-5mm}
\item[-~] the soft--breaking parameters $M_{\ti L}$,  $M_{\ti E}$, 
  $M_{\ti Q}$, $M_{\ti U}$, $M_{\ti D}$, $A_{\tau}$, $A_t$, $A_b$, 
  which determine the mass matrices of the stau, stop, and sbottom systems, 
  and the mass of the $\tau$-sneutrino, 
\item[-~] the (soft--breaking) $SU(2)$ and $U(1)$ gaugino masses $M$ and $M'$, 
\item[-~] the higgsino mass parameter $\mu$, and $\tan\beta = v_{2}/v_{1}$ 
  (where $v_1$ and $v_2$ are the vacuum expectation values of the neutral 
  members of the two Higgs doublets).  
\end{description}
We assume the GUT relations $M'/M = \frac{5}{3} \tan^2\Theta_W \approx 0.5$, 
and $m_{\ti g}/M = \alpha_s/\alpha_2 \approx 3$, where $m_{\ti g}$ is
the gluino mass. Furthermore, we assume that the $\nt_1$ is the lightest 
SUSY particle (LSP).

The lower model independent mass bound for stops obtained at LEP1
is 45 GeV \cite{lep1}.
Stronger limits up to 65 GeV are reported from the data taking at LEP1.5 at
130--161 GeV \cite{lep15}.
The D$\emptyset$ experiment at the TEVATRON excludes the
mass range $40$~GeV $\lets\:m_{\ti t}\:\lets\: 100$~GeV for the
stop, if the mass difference $m_{\ti t} - \mnt{1}~\grts~30$~GeV
\cite{D0}.

In Section~2 we shortly review the basic facts about L--R mixing
of stops, sbottoms, and staus. We also present numerical results for 
the production cross sections of stops, sbottoms, $\tau$-sneutrinos, 
and staus for unpolarized and polarized $e^-$ beams. 
In Section~3 we describe the decays of 
stops, sbottoms, $\tau$-sneutrinos, and staus and present
numerical results for the most important branching
ratios. We also list the signatures which are expected to be relevant
at $\sqrt{s} = 500$ GeV. In Section~4 we describe an event generator for
$\stst$ production and decay.
In Section~5 experimental sensitivities are determined based on
Monte Carlo simulations. In Section~6 we give an estimate of the experimental
error to be expected for stop and sbottom masses and the stop mixing angle, 
as well as for the soft SUSY breaking parameters. 
Section~7 gives a summary.

\section{Cross Sections for Pair Production of
         Stops, Sbottoms, \boldmath$\tau$-Sneutrinos, and Staus}

\subsection{Left-Right Sfermion Mixing}

The SUSY partners of the SM fermions with left and right
helicity are the left and right sfermions. In the case
of stop, sbottom, and stau the left and right states are in
general mixed. In the ($\sf_L,\,\sf_R$) basis
the mass matrix is \cite{Ellis,Gunion}
\beqa
M^2_{\tilde{f}} = \left( \begin{array}{cc}
                        m^2_{\tilde{f}_L} & a_f m_f \\
                        a_f m_f & m^2_{\tilde{f}_R}
                       \end{array}
                 \right)
\label{eq:sqmmat}
\eeqa
with
\beqa
  &&\hspace{-8mm} m^2_{\ti f_L} = 
  M^2_{\ti F} + m_Z^2 \cos 2\beta(T^3_f - e_f \sin^2\theta_W) + m^2_f , 
  \label{eq:msfl}\\
  &&\hspace{-8mm} m^2_{\ti f_R} = 
  M^2_{\ti F'} + e_f m_Z^2 \cos 2\beta \sin^2\theta_W + m^2_f , 
  \label{eq:msfr}\\
  &&\hspace{-8mm} a_t \equiv A_t - \mu \cot \beta, \hspace{2mm} 
  a_b \equiv A_b - \mu \tan \beta, \hspace{2mm} 
  a_{\tau} \equiv A_{\tau} - \mu \tan\beta,
  \label{eq:offdiag}
\eeqa
where $e_f$ and $T^3_f$ are the charge and the third component
of the weak isospin of the sfermion $\sf$, $M_{\ti F}=M_{\ti Q}$ 
for $\sf_L = \st_L,\, \sb_L$, $M_{\ti F}=M_{\ti L}$ for
$\sf_L = \stau_L$, $M_{\ti F'} = M_{\ti U},\, M_{\ti D},\,M_{\ti E}$ 
for $\sf_R = \st_R,\, \sb_R,\, \stau_R$,
respectively, and $m_f$ is the mass of the corresponding fermion.
From renormalization group equations \cite{dreema} 
one expects that due to the Yukawa interactions 
the soft SUSY breaking masses $M_{\ti Q}$, $M_{\ti U}$, $M_{\ti D}$, 
$M_{\ti L}$, and $M_{\ti E}$ of the $3^{rd}$ generation sfermions are 
smaller than those of the $1^{st}$ and $2^{nd}$ generation.
Evidently, $\st_L$-$\st_R$ mixing can be important because
of the large top quark mass. For sbottoms and staus L--R mixing can 
be important if $\tan \beta \;\grts\; 10$. The mass eigenvalues for
the sfermions $\sf = \st,\,\sb,\,\stau$  are
\beq
  m^2_{\ti f_{1,2}} = \onehf \left(m^2_{\ti f_L} + m^2_{\ti f_R}
  \mp \sqrt{(m^2_{\ti f_L} - m^2_{\ti f_R})^2 +
  4m_f^2 a^2_f} \,\right) \label{eq:msf12}
\eeq
where $\st_1$,  $\sb_1$ and $\stau_1$ denote the lighter
eigenstates. The mixing angles $\theta_{\ti f}$ are given by
\beq {\small 
  \cth_{\ti f} = 
  \frac{- a_f m_f}{\sqrt{(m_{\ti f_L}^2-m_{\ti f_1}^2)^2 + a_f^2 m_f^2}},
  \hspace{5mm} 
  \sth_{\ti f} = 
  \frac{m_{\ti f_L}^2-m_{\ti f_1}^2}
       {\sqrt{ (m_{\ti f_L}^2-m_{\ti f_1}^2)^2 + a_f^2 m_f^2}}.
} \label{eq:mixangl} \eeq
Hence, in the convention used we have 
$|\cth_{\ti f}| > \onesq\,$ if $\msf{L}<\msf{R}$ and 
$|\cth_{\ti f}| < \onesq\,$ if $\msf{R}<\msf{L}$. \\

\noi
The mass of the $\tau$-sneutrino is
\beq
  m^2_{\ti \nu_\tau} = M^2_{\ti L} + \onehf m^2_Z \cos 2\beta.
  \label{eq:snumass}
\eeq

\noi
The inversions of eqs. \eq{msfl} to \eq{snumass} are
\beqa
M^2_{\ti Q} &=& 
  m^2_{\ti t_2}\sin^2\theta_{\ti t} + m^2_{\ti t_1}\cos^2\theta_{\ti t} 
  - m^2_Z\cos 2\beta\left(\onehf - \twoth\sin^2 \theta_W\right) - m^2_t 
  \label{eq:mQst}\\
M^2_{\ti U} &=& 
  m^2_{\ti t_1}\sin^2\theta_{\ti t} + m^2_{\ti t_2}\cos^2\theta_{\ti t} 
  - \twoth\,m^2_Z\cos 2\beta\,\sin^2\theta_W - m^2_t \\
M^2_{\ti Q} &=& 
  m^2_{\ti b_2}\sin^2\theta_{\ti b} + m^2_{\ti b_1}\cos^2\theta_{\ti b} 
  + m^2_Z\cos 2\beta \left(\onehf - \oneth\sin^2\theta_W\right) - m^2_b 
  \label{eq:mQsb}\\
M^2_{\ti D} &=& 
  m^2_{\ti b_1} \sin^2\theta_{\ti b} + m^2_{\ti b_2}\cos^2\theta_{\ti b} 
  + \oneth\,m^2_Z\cos 2\beta\sin^2\theta_W - m^2_b \\
M^2_{\ti L} &=& 
  m^2_{\ti\tau_2}\sin^2\theta_{\ti\tau} + m^2_{\ti\tau_1}\cos^2\theta_{\ti\tau} 
  + m^2_Z\cos 2\beta \left(\onehf - \sin^2\theta_W\right) - m^2_\tau 
  \label{eq:mLstau}\\
M^2_{\ti E} &=& 
  m^2_{\ti \tau_1}\sin^2\theta_{\ti \tau} + 
  m^2_{\ti \tau_2}\cos^2\theta_{\ti \tau} + 
  m^2_Z \cos 2\beta \sin^2\theta_W - m^2_\tau \label{eq:mEstau}\\
M^2_{\ti L} &=& 
  m^2_{\ti \nu_\tau} - \onehf\, m^2_Z \cos 2\beta 
  \label{eq:mLsnu}\\
a_f m_f &=& 
  (m^2_{\ti f_1} - m^2_{\ti f_2}) \sin\theta_{\ti f}\cos \theta_{\ti f}
  \label{eq:afmf}
\eeqa
The soft--breaking parameter $M_{\ti Q}$ enters the equations 
of $\mst{L}$ and $\msb{L}$. Therefore eqs. \eq{mQst} and \eq{mQsb} 
imply the following condition (at tree--level):
\beq
  m^2_W \cos 2\beta = m^2_{\ti t_2} \sin^2 \theta_{\ti t} + m^2_{\ti t_1}
  \cos^2 \theta_{\ti t} - m^2_{\ti b_2} \sin^2 \theta_{\ti b} -
  m^2_{\ti b_1} \cos^2 \theta_{\ti b} - m^2_t + m^2_b.
  \label{eq:test}
\eeq
This condition shows that if $\tan\beta$ and five of the six measurable
parameters $m_{\ti t_1}, m_{\ti t_2}, \cos^2 \theta_{\ti t},
m_{\ti b_1}, m_{\ti b_2}$, and  $\cos^2 \theta_{\ti b}$ are
known, the sixth can be predicted. 
An analogous condition holds for $M_{\ti L}$ in the slepton sector 
due to eqs. \eq{mLstau} and \eq{mLsnu}.
Furthermore, from eqs. \eq{mixangl} and \eq{afmf} one can see that, 
in the convention used, $a_f$ and $\cos\theta_{\ti f}$ have opposite signs.

\subsection{Cross Sections for \boldmath$\eeto\sf_{i}\,\bar{\sf_{i}}$ }

The reaction $\eeto\sf_{i}\bar{\sf_{i}}$ proceeds via $\gamma$ and $Z$ 
exchange. The tree--level cross section at a center--of--mass energy 
$\sqrt{s}$ is given by \cite{Drees90,Hikasa,Bartl96}:

\beqa
  \s^{tree} = 
    \frac{\pi\a^2 N_C}{3s}\,\b^3 \left[\, Q_f^2 + \left(
    \frac{(v_e^2+a_e^2)\,v_{\ti f_i}^2}{16\,s^4_W c^4_W} \,s^{2}  
    - \right. \right. \hspace{5cm} & & \nonumber\\
  \left. \left.
  \frac{Q_f\,v_e\,v_{\ti f_i}}{2\,s^2_W c^2_W} \,s\,(s-m_Z^2) \right)
    \frac{1}{(s-m_Z^2)^2 + \Gamma_Z^2 m_Z^2} \,\right]\,    
    & & \label{eq:sigtree}
\eeqa

\noi
where $s^2_W = 1 - c^2_W = \sin^2\tW$, $v_e = 2\sin^2\tW - \onehf$, 
and $a_e = -\onehf$. 
$N_C$ is a colour factor which is 3 for squarks and 1 for sleptons. 
The tree--level cross section has the typical $\beta^{3}$ kinematic 
suppression where
$\b = \big(1 - 4\,\msf{i}^2/s \big) ^{1/2}$
is the velocity of the outgoing scalar particles.
The $Z$ coupling to $\sf_{i}\bar{\sf_{i}}$ is proportional $v_{\ti f_i}$
with $v_{\ti f_1} = 2\,(I^3_f\cos^2\t_{\ti f}-Q_f\sin^2\tW)$,
$v_{\ti f_2} = 2\,(I^3_f\sin^2\t_{\ti f}-Q_f\sin^2\tW)$,
where $I^3_f$ and $Q_{f}$ are the third component of the weak isospin
and the charge of the fermion $f$ ($Q_{e} = -1$). This coupling
vanishes at the mixing angles $\cos^2\theta_{\ti f} = 0.31,
0.16$, and $0.46$ for $\st_1$, $\sb_1$, and $\stau_1$,
respectively. Note that the sign of $\cos\theta_{\ti f}$ cannot
be determined from the cross section \eq{sigtree}, as this depends 
only on $\cos^{2}\theta_{\ti f}$.

The interference between
the $\gamma $ and $Z$ exchange contributions leads to a
characteristic minimum of the cross sections 
for $e^+ e^- \ra \sf_i \bar{\sf_i}$ which occurs at a specif\/ic 
value of the mixing angles $\theta_{\ti f}$ given by
\beq
  \cos^2\theta_{\ti f}|_{min} = \frac{e_f}{T^3_f} \sin^2\theta_W
  [1+(1-s/m^2_Z)  F(\sin^2 \theta_W)]\, 
\eeq
where
$F(\sin^2 \theta_W) = \cos^2 \theta_W (L_e + R_e)/(L_e^2 + R_e^2) 
\approx -0.22$, 
$F(\sin^2 \theta_W) = \cos^2 \theta_W /L_e \approx -2.9$, and  
$F(\sin^2 \theta_W) = \cos^2 \theta_W /R_e \approx 3.3$, 
for unpolarized, left-- and right--polarized $e^-$ beams, respectively, 
with $L_e = - \onehf + \sin^2\theta_W$ and $R_e = \sin^2 \theta_W$. 
For polarized $e^-$ beams the dependence on the mixing angles is much 
more pronounced than for unpolarized beams.
The corresponding minimum of the $\eeto\sf_{2}\bar{\sf_{2}}$ cross 
sections occurs at $1 - \cos^{2}\theta_{\ti f}|_{min}$.

In the calculations of the cross sections
we have also included SUSY--QCD corrections taking the formulae of
\cite{Eberl} (see also \cite{Drees90,Beenaker,djouadi}) 
with $\alpha_{s}(M_{Z}) = 0.12$, $m_{t} = 175$~GeV,
and corrections due
to initial state radiation \cite{Peskin}.

Figure~1\,a shows contour lines  of the total $\eeto\st_1\bar{\st_1}$ cross 
section in the $m_{\ti t_1} - \cos\theta_{\ti t}$ plane for 
$\sqrt{s} = 500$~GeV and unpolarized beams. 
For the calculation of the SUSY--QCD radiative corrections we have 
assumed $m_{\ti t_2}=m_{\ti g}=300$~GeV. 
Significantly above the threshold there is a clear dependence 
on $\cos\theta_{\ti t}$. 
Figure~1\,b shows the $\cos\theta_{\ti t}$ dependence of the 
$\eeto \st_1\bar{\st_1}$ cross section for left-- and 
right--polarized as well as for unpolarized $e^-$ beams for $\sqrt{s} =
500$~GeV and  $m_{\ti t_1} = 180$~GeV. 
For both left-- and right--polarized $e^-$ beams the cross
sections depend strongly on the mixing angle. It is important to
note that this dependence is opposite for left and
right polarization. Therefore,
experiments with polarized $e^-$ beams will allow a more precise 
determination of the mass $m_{\ti t_1}$ and the mixing angle 
$\theta_{\ti t}$.

Figure 2\,a shows contour lines of the total cross section
of $\eeto \st_2 \bar{\st_2}$ in the $m_{\ti t_2} - \cos^2\theta_{\ti t}$ 
plane at $\sqrt{s} = 2$~TeV. 
The $\cos\theta_{\ti t}$ dependence of this cross section at 
$\sqrt{s} = 2$~TeV for left-- and right--polarized and unpolarized 
$e^-$ beams is shown in Fig.~2\,b for $m_{\ti t_2}=700$~GeV. 
Again, the $\cos\theta_{\ti t}$ dependence is much stronger for polarized
than for unpolarized beams, however, the behavior is opposite to that of
$\eeto \st_1 \bar{\st_1}$. 
For the calculation of the SUSY--QCD radiative corrections we assumed 
$m_{\ti t_1}=300$~GeV and $m_{\ti g}=700$~GeV.

We shall discuss in Section~6 how experimental data on cross sections 
with left-- and right--polarized beams would allow to determine 
masses and mixing angles, and then give information on soft SUSY 
breaking parameters.
 

Figure~3\,a shows the contour plot of the total cross section of 
$\eeto \sb_1 \bar{\sb_1}$ in the $m_{\ti b_1} - \cos\theta_{\ti b}$ 
plane at $\sqrt{s} = 500$~GeV for unpolarized beams. 
For a polarized $e^-$ beam the $\cos\theta_{\ti b}$ dependence of the cross 
sections is much stronger, as shown in Fig.~3\,b for $m_{\ti b_1} = 180$~GeV 
and  $\sqrt{s} = 500$~GeV.

Contour lines for the cross section of $\eeto \stau_1\bar{\stau_1}$ 
in the $m_{\ti \tau_1} - \cos\theta_{\ti \tau}$ plane
at $\sqrt{s}=500$~GeV are shown in Fig. 4\,a. 
Figure 4\,b shows the cross section for polarized and unpolarized $e^-$ beams 
as a function of $\cos\theta_{\ti\tau}$ for $m_{\ti \tau_1}=180$~GeV and 
$\sqrt{s}=500$~GeV. For both beam polarizations these cross sections again 
exhibit a strong dependence on the mixing angle.

Figure 5 shows the $m_{\ti \nu}$ dependence of the cross
section for $\eeto \snu_\tau \bar {\snu_\tau}$ for
unpolarized as well as left-- and right--polarized $e^-$ beams.
Since the $\snu_\tau$ is not mixed, the polarization dependence is 
entirely due to the different $Z e^+ e^-$ couplings. 

The $\sqrt{s}$ dependence of the
 $\eeto \st_1 \bar{\st_1}$ cross section is shown in Fig. 6
for $m_{\ti t_1}=180$~GeV and $\cos\theta_{\ti t}=0.7$. 
The effect of SUSY--QCD corrections for $\eeto \st_1 \bar{\st_1}$ from 
gluon and gluino exchange as well as the initial state radiation
correction as a function of $\sqrt{s}$ is demonstrated in Fig. 7, 
for $\cos\theta_{\ti t} = 0.7$, $m_{\ti t_1} = 180$~GeV, and
$m_{\ti t_2} = m_{\ti g} = 300$~GeV.
Note that at high energies the gluino exchange contribution
has the opposite sign of the gluon exchange contribution, 
and the absolute values are increasing with $\sqrt{s}$.
The effects are similar for $\eeto \sbsb$.
A detailed discussion of SUSY--QCD corrections is given in \cite{Eberl}. 
The effect due to initial state radiation turns out to be 
of the order of 10 \%. The sum of all corrections can well exceed 10\%.

\section{Decays of Stop, Sbottom, \boldmath$\tau$-Sneutrino, and Stau}

The sfermions of the third generation can have the weak decays 
($i,\,j = 1,\,2$; $k = 1,\ldots,\,4$ )
\beqa
  \st_i       &\ra& t\,\nt_k,    \hspace{4mm} b\,\chp_j  \label{eq:stdec}\\
  \sb_i       &\ra& b\,\nt_k,    \hspace{4mm} t\,\chm_j  \label{eq:sbdec}\\
  \stau_i     &\ra& \tau\,\nt_k, \hspace{3.5mm} \nu_{\tau}\,\chp_j 
    \label{eq:staudec}\\
  \snu_{\tau} &\ra& \nu\,\nt_k,  \hspace{3.5mm} \tau\,\chm_j 
    \label{eq:snudec}
\eeqa
Owing to the Yukawa terms and the L--R mixing the decay patterns of 
stops, sbottoms, and staus will be different from those of the sfermions of 
the first two generations \cite{baer, Bartl91}.
Stops and sbottoms can also have the strong decays
\beq
  \st_i \ra t\,\sg, \qquad \sb_i \ra b\,\sg.
    \label{eq:sqglu}
\eeq
They are dominant if they are kinematically allowed.
Otherwise, the lighter squark mass eigenstates decay mostly according 
to \eq{stdec} and \eq{sbdec}.
Moreover, in case of strong L--R mixing the splitting
between the two mass eigenstates may be so large that the
following additional decay modes are present \cite{Bartl94}:
\beqa
 \st_2 &\ra& \st_1\,Z\, (h^0,\, H^0,\, A^0), \qquad \sb_1\,W^+\, (H^+), 
    \label{eq:st2dec}\\
 \sb_2 &\ra& \sb_1\,Z\, (h^0,\, H^0,\, A^0), \qquad \st_1\,W^-\, (H^-),
    \label{eq:sb2dec}\\
 \stau_2 &\ra& \stau_1\,Z, \hspace{3.1cm} \snu_\tau\,W^-.
    \label{eq:stau2dec}
\eeqa

\noi 
If the $\st_1$ is the lightest charged SUSY particle and 
$\mnt{1} + m_b + m_W < m_{\ti t_1} < \mnt{1} + m_t$,
the decay $\st_1 \ra b \, W^+ \, \nt_1$
can be dominant \cite{Porod}, otherwise the higher--order decay
$\st_1 \ra c \, \nt_1$ dominates \cite{Hikasa}. 
In the case that $m_{\ti \tau_1} < m_{\ti t_1}$ also
$\st_{1}\ra b\,\nu_{\tau}\,\stau_{1}$ and in case that 
$m_{\ti \nu} < m_{\ti t_1}$ also $\st_{1}\ra b\,\tau^{+}\,\snu_{\tau}$ 
might be important.

Figure 8\,a and 8\,b show the parameter domains in the $M-\mu$ plane for the 
decays of $\st_1$ and $\sb_1$, eqs. \eq{stdec} and \eq{sbdec}, taking
$m_{\ti t_1} = 180$~GeV, $\tan \beta = 2$, and $m_{\ti b_1} = 180$~GeV, 
$\tan \beta = 30$. 
In region (a) only the decay $\st_1 \ra c\, \nt_1$ is allowed, whereas 
in region (b) $\st_1 \ra c\, \nt_1$ and $\st_1 \ra b\, W^+ \nt_1$ are 
possible. 
In the small stripe of region (c) also $\st_1 \ra c\,\nt_2$ is possible. 
In region (d) the decay $\st_1 \ra b\,\chp_1$ has practically 100\% branching 
ratio. 
This is further illustrated in Fig. 9\,a where we plot the branching
ratios for $\st_1$ decays as a function of $M$ for $\mu = -500$~GeV, 
$m_{\ti t_1} = 180$~GeV, $\tan \beta = 2$, and $\cos\theta_{\ti t} = 0.7$.
The parameter domains for the
$\stau_1$ decays into neutralinos 
are almost identical to those of
the corresponding $\sb_1$ decays, if the masses of
 $\stau_1$ and $\sb_1$ are equal. 
 
Figure~9\,b shows the branching ratios for $\sb_1$ decays as a function of $M$ 
for $m_{\ti b_1} = 180$~GeV, $\tan \beta = 30$, $\cos\theta_{\ti t} = 0.7$, 
and $\mu = -500$~GeV. 
Figure 10\,a and 10\,b show the branching ratios for $\sb_1$ and $\stau_1$ 
decays as a function of $M$ for $\mu = - 130$~GeV and  
$m_{\ti b_1} = m_{\ti \tau_1} = 180$~GeV, $\tan \beta = 30$, 
$\cos\theta_{\ti b} = \cos\theta_{\ti \tau} = 0.7$. 
Similarly, Figure~10\,c shows the branching ratios for $\snu_{\tau}$ decays  
taking $m_{\ti \nu_{\tau}} = 180$~GeV, $\tan \beta = 2$, and
$\mu = - 130$~GeV. 
The LSP decays of $\sb_1$ and $\stau_1$ are about $60\%$ and $40\%$, 
respectively. The branching ratio for the visible decays of the 
$\snu_{\tau}$ is between $40\%$ and $90\%$.

The decay patterns of the heavier sfermion mass--eigenstates can be
quite complicated, because all the decay modes of eqs. \eq{stdec} to 
\eq{stau2dec} can occur. 
We calculate the different decay widths with the formulae of 
Refs. \cite{Bartl94,Bartl96,Porod}. 
Figure 11\,a and 11\,b show the branching ratios for $\st_2$
decays as a function of $M$ for $m_{\ti t_2}=700$~GeV, taking
$\tan\beta=2$, $\mu =-1000$~GeV, $m_A=150$~GeV, $M_{\ti
Q}=607$~GeV, $M_{\ti U}=360$~GeV,  $M_{\ti D}=850$~GeV,
and $A_t=500$~GeV. 
The masses and mixing parameters of the other particles involved 
then are $m_{\ti t_1}=258$~GeV, $\cos\theta_{\ti t}=-0.468$, 
$m_{\ti b_1}=608$~GeV, $m_{\ti b_2}=851$~GeV, $\cos\theta_{\ti b}=-0.999$,
$m_{h^0}=98$~GeV, $m_{H^0}=165$~GeV, $m_{H^{\pm}}=169$~GeV,
$\cos\alpha =0.535$. 
We have included radiative corrections to the Higgs masses and the 
Higgs mixing angle according to \cite{higgscorr}.
For $M \,\lets\, 200$~GeV the decay $\st_2 \ra
t \sg$ dominates, whereas for $M \,\grts\, 200$~GeV the decay 
$\st_2 \ra Z \st_1$ has the largest branching fraction. 
It is interesting to note that in this case the decays 
$\st_2 \ra A^0\,\st_1$ and $\st_2 \ra H^0\, \st_1$ have larger 
branching ratios than the decay into $b\, \chp_1$ which is the most
important decay mode of $\st_1$. 

\noi
Figure 12\,a and 12\,b show
the branching ratios of the $\sb_2$ decays for $m_{\ti b_2}=700$~GeV, 
$\tan\beta=30$, $\mu =-1000$~GeV, $m_A=150$~GeV, 
$M_{\ti Q}=637$~GeV, $M_{\ti U}=320$~GeV, $M_{\ti D}=450$~GeV, 
and $A_t=500$~GeV. 
The masses and mixing parameters of the other particles involved 
then are 
$m_{\ti b_1}=350$~GeV, $\cos\theta_{\ti b}=-0.469$, $m_{\ti t_1}=324$~GeV, 
$m_{\ti t_2}=678$~GeV, $\cos\theta_{\ti t}=-0.273$, $m_{h^0}=118$~GeV, 
$m_{H^0}=146$~GeV, $m_{H^{\pm}}=176$~GeV, and $\cos\alpha =0.998$. 
In this example the transition $\sb_2 \ra H^-\, \st_1$ is dominant for 
$M \,\grts\, 200$~GeV, and $\sb_2 \ra Z\, \sb_1$ and $\sb_2 \ra W^-\,\st_1$ 
have larger branching ratios than $\sb_2 \ra t\, \chp_1$ or
$\sb_2 \ra b\, \nt_{1,2}$. 

\noi
The branching ratios for the decays of the sleptons $\stau_2$ and 
$\snu_{\tau}$ are shown in Figs. 13 a, b, and Figs. 14 a, b, for 
$m_{\ti\tau_2}=700$~GeV and $m_{\ti \nu_{\tau}}= 687$~GeV,
respectively, taking  $\tan\beta=30$, $\mu =-1000$~GeV, $m_A=150$~GeV, 
$M_{\ti L}=690$~GeV, $M_{\ti E}=490$~GeV, and $A_{\tau}=500$~GeV. 
The mass of the lighter stau then is $m_{\ti \tau_1}=480$~GeV, and 
the mixing angle is $\cos\theta_{\ti \tau}=-0.213$.
The Higgs boson masses are the same as in Fig. 12. 
In these examples, the $\stau_2$ and $\snu_{\tau}$ decays into charginos 
and neutralinos have large branching ratios, and the decays 
$\stau_2 \ra \nu\,\chm_1$ and $\snu_{\tau} \ra \tau\,\chp_1$ dominate.
The decay $\stau_2 \ra Z\,\stau_1$ has a branching ratio of
$5\%$ to $10\%$, 
and $\snu_{\tau} \ra \tau\,\chp_1$ has a branching ratio of 
$35\%$ to $55\%$. 
It is interesting to note that in this example the branching ratio of 
the invisible decays mode $\snu_{\tau} \ra \nu_{\tau}\, \nt_1$ is always 
less then about $18\%$.

Quite generally, the decay widths of the $\st_{i}$, $\sb_{i}$, and 
$\stau_{i}$ decays into neutralinos and charginos, eq.~\eq{stdec} to 
\eq{staudec}, depend also on the sign of $\cos\theta_{\ti f}$.
It may therefore be possible to determine the sign of the mixing 
angle by studying these decays.

Table~1 lists the most important signatures for $\st_{1}$, $\sb_{1}$, 
$\snu_{\ti\tau}$, and $\stau_{1}$ for $\sqrt{s} = 500$~GeV.
If the decays $\st_{1}\ra b\,\chp_{1}$ or  
$\stau_{1}\ra \nu_{\tau}\,\chm_{1}$ occur, the $\ch_{1}$ will most 
probably be discovered first and thus its mass and couplings
will be known. 
The decay $\st_{1}\ra b\, W^{+}\, \nt_{1}$ leads to the same
final states as $\st_{1}\ra b\, \chp_{1}$ (provided 
$\chp_{1} \ra H^+ \, \nt_{1}$ is not allowed).
The decay $\snu_{\tau}\to\nu\nt_{1}$ is invisible. Thus, one--sided 
events can occur where one $\snu_{\tau}$ decays invisibly and the 
other one decays visibly into one of the f\/inal states given in table~1.


\vspace{3mm}

\begin{center}
\begin{tabular}{|l|l|}   
\hline 
 Channel  &  Signatures  \\
\hline  
  $\st_{1}\to b\,\chp_{1}$ & 1 $b$-jet + 1 $\ell^{+}$ + $\ptmiss$,
                             1 $b$-jet + 2 jets + $\ptmiss$ \\
\hline 
  $\st_{1}\to c\,\nt_{1}$  & 1 jet + $\ptmiss$ \\
\hline 
  $\sb_{1}\to b\,\nt_{1}$  & 1 $b$-jet + $\ptmiss$ \\
\hline
  $\sb_{1}\to b\,\nt_{2}$  & 1 $b$-jet + $\ell^{+}\ell^{-}$ + $\ptmiss$,
                             1 $b$-jet + 2 jets + $\ptmiss$ \\
\hline 
  $\stau_{1}\to \tau\,\nt_{1}$ & $\tau + \ptmiss$ \\
\hline
  $\stau_{1}\to \tau\,\nt_{2}$ & $\tau + \ell^{+}\ell^{-} + \ptmiss$,
                                 $\tau$ + 2 jets + $\ptmiss$ \\
\hline
  $\stau_{1}^{-}\to \nu_{\tau}\,\chm_{1}$ & $\ell^{-} + \ptmiss$,
                                        2 jets + $\ptmiss$ \\
\hline
  $\snu_{\tau}\to \tau^{-}\,\chp_{1}$ & $\tau^{-} + \ell^{+} + \ptmiss$,
                                    $\tau^{-}$ + 2 jets  + $\ptmiss$ \\
\hline
  $\snu_{\tau}\to \nu\,\nt_{2}$   & $\ell^{+}\ell^{-} + \ptmiss$,
                                    2 jets  + $\ptmiss$, 
                                    $\gamma$ + $\ptmiss$ \\
\hline  
\end{tabular} 
\refstepcounter{table}   
\label{tab:signatures}
\vspace{2mm}

\noi
\begin{tabular}{lp{11cm}}   
  {\bf Table~\arabic{table}:} & 
  Expected signatures for $\st_{1}$, $\sb_{1}$, $\snu_{\tau}$, 
  and $\stau_{1}$ production for $\sqrt{s} =$~500~GeV.  
  Owing to pair production all combinations of the corresponding  
  signatures may occur.
\end{tabular} 

\end{center}

\section{Stop Event Generation}

In this section we describe the event generator for $\eeto\st_1\bar{\st}_1$
with the stop decay modes $\st_1 \ra c \chiz$ and $\st_1 \ra b \chip$.
The chargino decays via $\chip \ra W^{+} \chiz$, where $W^{+}$
can be either virtual or real.
The event generator is based on the calculation
of the 4-momenta distributions of the stop decay products
\mbox{\chiz $c$ \chiz $\bar{c}$} and \mbox{\chip $b$ \chim 
$\bar{b}$}.
The large effects of QCD corrections are included in the
cross section calculation.
Stop production and decay have been defined as new processes in the 
PYTHIA program package~\cite{pythia}.
The event generation process includes the modelling of hadronic
final states. 

In the first step of the event generation, initial state photons are
emitted using the program package REMT~\cite{pythia} which takes into 
account the expected stop cross section from zero to the nominal 
center-of-mass energy. Beamstrahlung photons are generated using the 
beam parameters of the NLC 1992 design.
The effective center-of-mass energy is calculated for the
initial production of the 4-momenta of the final-state particles.
These 4-momenta are then boosted to the lab-frame 
according to the momentum of the emitted photons.
For the hadronization process of the \cc\ in the 
\mbox{\chiz $c$ \chiz $\bar{c}$}
 and
of the \bb\ in the \mbox{\chip $b$ \chim $\bar{b}$} decay mode, 
a color string
with invariant mass of the quark-antiquark-system is defined. The possible
gluon emission and hadronization are performed using the Lund model
of string fragmentation with the PYTHIA program package~\cite{pythia}.
The Peterson {\it et \,al.}~\cite{peterson} fragmentation parameters 
for the $c$- and $b$-quarks are used: 
$\epsilon_c=0.03$ and $\epsilon_b=0.0035$.
Finally, short-lived particles decay into their observable final
state.
Details of the event generator are given in~\cite{sopcz}.

\section{Simulation and Selection}

The investigated background reactions and their cross sections
are shown in Fig.~15. They are simulated for \lumifb{10},
and 1000 signal events are simulated in
the \chichi\ and \chipchip\ decay channels. 
The L3 detector at CERN including the upgrades for LEP2
served as an example for an \ee\ \gev{500} detector. Details of the
parametric detector simulation are given in~\cite{zphys}.
An important feature is the overall hadronic energy resolution of
about 7\%. 

In both channels, 
the \chiz 's escape the detector and cause large missing energy.
In the case of \chichi, the $c$-quarks form mostly two acoplanar jets. 
A mass combination of $M_{\st_1}=\gev{180}$ and $\mnt{1}=\gev{100}$
is investigated in detail. For {\chipchip} on average the
visible energy is larger. In this channel, the mass combination
$M_{\st_1}=\gev{180}$, 
$m_{\chp_1}=\gev{150}$, and $\mnt{1}=\gev{60}$ has been studied.
Typically four jets are formed, two from the $b$-quarks,
and two from the boosted $W$'s.

In the first step of the event selection, unbalanced hadronic events
are selected using the following selection requirements:
$$
25 <  {\mathrm{hadronic~clusters}} < 110,~~~~~
0.2 < E_{vis}/\sqrt{s} < 0.7,
$$
$$
E^{\mathrm{imb}}_{\parallel}/E_{\mathrm{vis}} < 0.5,~~~~~
\mathrm{Thrust} < 0.95,~~~~~
\abs{\cos\theta_{\mathrm{Thrust}}}<0.7~.
$$

\vspace{3mm}

\begin{center}
\begin{tabular}{|c|c|c|c|c|c|c|c|c|}\hline
Channel &\chichi&\chipchip& qq & WW & eW$\nu$& tt &ZZ  & eeZ \\ \hline
Total (in 1000)&1& 1      &125 & 70 &  50    &  7 & 6  & 60    \\ \hline
After preselection (in 1000) & 0.4   & 0.7   & 1.7& 2.2&  3.2   & 1.3& 0.2& 0.3   \\ \hline
\end{tabular}
\vspace{2mm}
\refstepcounter{table}   
\label{tab:pres}

\noi
\begin{tabular}{lp{12cm}}   
  {\bf Table~\arabic{table}:} & 
  Expected events per 10~f\/b$^{-1}$ at $\sqrt{s} = \gev{500}$, 
  and number of events after the preselection as defined in the text.
\end{tabular} 
\end{center}

\noi
A large part of the background of back-to-back events without missing
energy is rejected.
Table~\ref{tab:pres} shows the number of initially produced events per
${\cal L}$ = 10 $\rm f\/b^{-1}$ at $\sqrt{s}=500$~GeV, and the number 
of events which pass this preselection. 
The requirement of a large number of hadronic clusters
removes \ee, \mm, and most of the \tautau\ events. 
The minimum energy cut reduces
most of the $\gamma\gamma$ events and ensures almost
100\% trigger efficiency.
The background from $\gamma\gamma$ events can, in addition, be strongly 
reduced by rejecting events where a scattered initial electron is detected
at low angles.
The upper energy cut reduces all standard background reactions.
Beam gas events and events where much energy goes undetected along the beam
axis are removed by rejection of events with very large
parallel imbalance. 
The thrust cut removes remaining \tautau\ events and reduces 
largely \qq\ and \ZZ\ background.
The $\cos\theta_{\mathrm{Thrust}}$ cut removes events where most 
probably much energy escapes 
undetected along the beam axis.

The final \chichi\ event selection is summarized in Table~\ref{tab:final}.
The following cuts are applied: 
\begin{itemize}
\vspace*{-3mm}
\item A hard upper energy cut reduces all standard background
      except $e W \nu$  (Fig.~16).
\vspace*{-8mm}
\item Jets are clustered using the JADE algorithm.
      The y-cut value is optimized to obtain two jets for the signal.
\vspace*{-3mm}
\item Semileptonic decays of the top quark can induce missing energy.
      These events are partly removed by requiring no isolated electron 
      or muon.
\vspace*{-3mm}
\item Events with large longitudinal energy imbalance are removed 
      where probably much energy escapes undetected along the beam axis.
\vspace*{-3mm}
\item The invariant mass of the two jets is required to be larger
      than \gev{120} to remove almost entirely $eW\nu$ events
      (Fig.~17).
\vspace*{-3mm}
\item The acoplanarity angle is defined as the angle between 
      the jets in the plane perpendicular to the beam axis. 
      A maximum value of 2.9~rad is important to reduce the remaining 
      background. \vspace*{-2mm}
\end{itemize}

The result of this study is
4.3\% detection efficiency and 9 background events.
A detection confidence level of 3$\sigma$ (99.73\%) is expected 
for a cross section of 23~f\/b. Expected signal and background
are shown in Fig.~18.

\vspace*{3mm}

\begin{center}
\begin{tabular}{|c|c|c|c|c|c|c|c|}\hline
Channel &$\nt_1$c$\nt_1 \bar{c}$& qq & WW &eW$\nu$& tt &ZZ  & eeZ   \\ \hline
Total (in 1000)    & 1     &125 & 70 &  50 &  7 & 6  & 60    \\ \hline

After Preselection & 391   &1652&2163&3185 &1259&182 & 318   \\ \hline

$E_{vis}/\sqrt{s}<0.4$  
                  & 332   &{\bf 202}&{\bf 285}&3032   &{\bf 70}&{\bf 4}&{\bf 98} \\ \hline

Njet $=2$          & 293   & 172& 182&2892   &{\bf 17}&  3 &  72 \\ \hline

No isolated e or $\mu$
                   & 218   & 152&{\bf 98}&2757&{\bf 5}&  3 &{\bf 9} \\ \hline

$E^{imb}_{\parallel}/E_{vis}<0.3$
                   &185 &{\bf 101}&  70&2049   &   5&  2 &   4   \\ \hline

Invariant\,mass\,of\,jets$>$120GeV
                   & 52    &{\bf 25}&{\bf 12}&{\bf 7}&1&  0 &   0 \\ \hline
Acoplanarity $<2.9$rad&43     &{\bf 0} &{\bf 5}&{\bf 3}&  1&  0 &   0  \\ \hline
\end{tabular}
\vspace{2mm}
\refstepcounter{table}   
\label{tab:final}

\noi
\begin{tabular}{lp{11cm}}   
  {\bf Table~\arabic{table}:} & 
  Final event selection cuts, expected signal efficiencies, and the  
  number of expected background events. Bold face numbers indicate  
  major background reductions.
\end{tabular} 
\end{center}

The final \chipchip\ event selection is summarized in Table~\ref{tab:final2}.
Here the cuts are:
\begin{itemize}
\sloppy
\vspace*{-3mm}
\item A hard lower energy cut reduces most of the $eW\nu$ background.
\vspace*{-3mm}
\item Topologies with back-to-back jets are reduced by an upper cut on
      the event thrust (Fig.~19).\vspace*{-3mm}
\item A lower cut on the number of hadronic clusters reduces efficiently
      low-multiplicity background final states
      (Fig.~20).\vspace*{-3mm}
\item Jets are clustered using the JADE algorithm.
      The y-cut value is optimized to obtain four jets for the signal.
\vspace*{-3mm}
\item Events with an isolated electron or muon are rejected.
\vspace*{-3mm}
\item An upper cut on the visible energy reduces \qq, \WW, and 
      $t \bar{t}$ background.
\vspace*{-3mm}
\item Finally, the remaining $t \bar{t}$ background events are 
reduced by requiring less than 30\% perpendicular energy 
imbalance.\vspace*{-3mm}
\end{itemize}

Concerning the number of $b$-quarks per event, the decay
$\chipchip \ra W^+\chiz bW^-\chiz \bar{b}$ leads to the same final 
states as expected for $t \bar{t}$ background. 
Therefore, the tagging of $b$-quarks has not 
proved to be efficient to reduce this background.

The result of this study is
4.5\% detection efficiency and 8 background events.
A detection confidence level of 3$\sigma$ (99.73\%) is expected 
for a cross section of 19~f\/b.
Expected signal and background are shown in 
Fig.~21.

\vspace*{3mm}

\begin{center}
\begin{tabular}{|c|c|c|c|c|c|c|c|}\hline
Channel            &$\chp_1 b\chm_1\bar{b}$& qq & WW &eW$\nu$& tt &ZZ  & eeZ   \\ \hline
Total (in 1000)    & 1    &125 & 70 &  50   &  7 & 6  & 60    \\ \hline

After Preselection & 695  &1652&2163&3185   &1259&182 & 318   \\ \hline

$E_{vis}/\sqrt{s}>0.35$ 
                   & 610  &1494&2011&{\bf 337}&1234&178 & 239 \\ \hline

Thrust $<0.85$     & 536  &{\bf 326}&{\bf 420}&{\bf 24}&1141&{\bf 69}&137 \\ \hline   

Ncluster $\geq 60$ & 399  & 195&{\bf 134}&{\bf 0}  & 769& 41 &{\bf 3}     \\ \hline 

Njet $=4$          & 211  &{\bf 53}&{\bf 72}& 0  & 432& 22 &   0    \\ \hline

No isolated e or $\mu$& 99&  41&  49&   0  &{\bf 105}& 16 & 0 \\ \hline

$E_{vis}/\sqrt{s}<0.55$&57&{\bf 3}&{\bf 8}& 0 &{\bf  23}& 0 & 0 \\ \hline

$E^{imb}_{\perp}/E_{vis}<0.3$& 45     &  1& 3& 0 & {\bf 4}&  0 & 0 \\ \hline

\end{tabular}
\vspace{2mm}
\refstepcounter{table}   
\label{tab:final2}

\noi
\begin{tabular}{lp{11cm}}   
{\bf Table~\arabic{table}:} & Final event selection cuts, expected 
signal efficiencies, and the number of expected background events.  
Bold face numbers indicate major background reductions.
\end{tabular} 
\end{center}

At a future \ee\ collider with $\sqrts=\gev{500}$, a large discovery
potential for scalar top quarks is already expected within one
year of data-taking (\lumifb{10}).
Detector performances known from LEP detectors result in good
background reduction. Full hermeticity of the detector is essential.

The confidence levels for discovering a signal are shown in
Figs.~22\,a and 22\,b for the $\chipchip$ and $\chichi$ channels,
respectively. Here, the confidence levels are given in
$\sigma = N_{\mathrm{expected}} / \sqrt{N_{\mathrm{background}}}$.
The sensitivity is sufficient to discover a \gev{200} stop 
independently of the values of the mixing angle
with $3\sigma$ in both \chiz $c$ and \chip $b$ decay modes 
for the investigated neutralino and chargino mass
combinations. A complete set of mass combinations remains to be
studied. 

At a later stage, the total luminosity could reach 50~$\fb^{-1}$, and the 
resulting discovery region for a $3\,\sigma$ effect is shown in 
Fig.~23. An increase of the center--of--mass energy would extend the 
discovery region further as shown in Fig.~24 for $\sqrt{s} = 800$~GeV 
${\cal L} = 200~\fb^{-1}$. 
Based on experience made in the LEP2 searches \cite{L3note} the 
efficiency for the simulated mass combination can be extended to a 
larger mass region for $\mst{1}-\mch{1} > 20$~GeV.

\clearpage 
\section{Determination of Soft--Breaking Parameters --- A Case Study}

In this section we want to estimate the experimental accuracies
for the stop and sbottom masses and mixing angles
which can be expected from the Monte Carlo simulation
described in the preceding sections.
Without beam polarization a possible way to determine
$\mst{1}$ and $\cos\theta_{\ti t}$ is using the
$\sqrt{s}$ and $\cos\theta_{\ti t}$ dependence
of the unpolarized $e^+e^- \ra \st_1 \bar {\st_1}$
total cross section (see Figs. 1a and 6). 
Let us take as reference point $\mst{1}=180$~GeV,
$\cos\theta_{\ti t}=0.57$, and $\sqrt{s}=400$~GeV and
$\sqrt{s}=500$~GeV as the two reference energies.
Note that at $|\cos\theta_{\ti t}=0.57|$ the 
$e^+e^- \ra \st_1 \bar {\st_1}$ cross section has its minimum. 
The cross sections at this point for these two energy values are
$\sigma = 18.2 \pm 4.1$~f\/b at $\sqrt{s}=400$~GeV and 
$\sigma = 47.4 \pm 5.5$~f\/b at $\sqrt{s}=500$~GeV 
where the experimental errors follow from the Monte Carlo
simulation.
Figure~25 shows the corresponding error bands in the
$\mst{1} - \cos\theta_{\ti t}$ plane. 
As can be seen, hardly an information can be obtained on 
the mixing angle. 

The polarization of the $e^{-}$ beam offers the possibility of 
measuring the sfermion masses and especially the mixing angles with 
much higher accuracy.
The cross sections of $e^+e^- \ra \st_1 \bar {\st_1}$ for 90\% left-- 
and right--polarized $e^-$ beam at the reference point $\mst{1}=180$~GeV, 
$|\!\cos\theta_{\ti t}| = 0.57$ for $\sqrt{s} = 500$~GeV are
$\sigma_L = 48.6 \pm 6.0\;{\rm f\/b}$,  
$\sigma_R = 46.1 \pm 4.9\;{\rm f\/b}$,
where the experimental errors are given by  
$\Delta\sigma / \sigma = 
 N_{signal} \,/\, \sqrt{N_{signal}+N_{background}}$
with the number of signal and background events determined as 
described in the previous section.
Figure~26 shows the correponding error bands and the error ellipse 
in the $\mst{_1} - \cos\theta_{\ti t}$ plane. 
The experimental accuracies obtained in this way for the mass of 
the lighter stop and the stop mixing angle are 
\beqa
  m_{\ti t_1}&=&180 \pm 7\; {\rm GeV}, \\
  \cos\theta_{\ti t}&=&0.57 \pm 0.06.
\eeqa

We treat the sbottom system in an analogous way. 
Assuming that $\tan\beta$ is not too large we can neglect left--right 
mixing in the sbottom sector.
In the ``Minimal Supergravity--inspired Model'' \cite{msugra} one expects 
$\msb{L}\,\lets\,\msb{R}$, thus $\sb_1 = \sb_L$ and $\sb_2 = \sb_R$, 
i.e. $\cos\theta_{\ti b} = 1$. 
As reference point of the sbottom system we take $m_{\ti b_1} = 200$~GeV, 
$m_{\ti b_2} = 220$~GeV. 
The cross sections for $e^+e^- \ra \sb_1 \bar {\sb_1}$ with 90\% 
left--polarized $e^-$ beams and for $e^+e^- \ra \sb_2 \bar{\sb_2}$ 
with 90\% right--polarized $e^-$ beams then are
$\sigma_L(e^+e^- \ra \sb_1 \bar{\sb_1}) = 61.1 \pm 6.4$~f\/b, 
$\sigma_R(e^+e^- \ra \sb_2 \bar{\sb_2}) = 6.0 \pm 2.6$~f\/b,
where the errors are again determined by our Monte Carlo procedure. 
The errors for the sbottom masses follow as:
\beqa
  m_{\ti b_1}&=&200 \pm ~\,4\;  {\rm GeV}, \\
  m_{\ti b_2}&=&220 \pm 10\; {\rm GeV}.
\eeqa

\noi
With these values for $m_{\ti t_1}$, $\cos\theta_{\ti t}$, 
$m_{\ti b_1}$, and $m_{\ti b_2}$ we can use \eq{test} 
and obtain the mass of the heavier stop $\st_{2}$ if $\tan\b$ is 
known from other experiments. Taking, for instance, $\tan\b = 2$
leads to
\beq
  m_{\ti t_2} = 289 \pm 15\; {\rm GeV}.
\eeq
Confirming this value by producing $\st_{2}\bar{\st_{2}}$ at higher
energies would be an independent test of the MSSM. 

Assuming that also $\mu$ is known from an other experiment 
we are now able to calculate the underlying soft SUSY breaking 
parameters $M_{\ti Q}$, $M_{\ti U}$, $M_{\ti D}$, $A_{t}$ and 
$A_{b}$ for the  squarks of the third family according to 
eqs.~(8) to (15). 
Taking $\mu = - 200$~GeV, $\tan\beta=2$, and $m_t=175$~GeV we 
obtain the following values:
\beqa
  M_{\ti Q}&=&195 \pm 4\;  {\rm GeV},   \\
  M_{\ti U}&=&138 \pm 26\; {\rm GeV},  \\
  M_{\ti D}&=&219 \pm 10\; {\rm GeV},  \\
  A_t &=& -236 \pm 38\; {\rm GeV} \qquad {\rm if}\;\cos\theta_{\ti t} > 0, \\
  A_t &=& 36 \pm 38\;{\rm GeV} \qquad 
    {\rm if}\; \cos\theta_{\ti t} < 0.
\eeqa

These results have to be compared with those of \cite{nlc} and 
\cite{daniel}, where the stop mass was determined by a 
kinematical reconstruction of the $b\chp_{1}$ decay. 
In \cite{finel} a somewhat higher accuracy was obtained for unmixed 
squarks because one parameter less is involved in this case.


\section{Summary}

In this article we have discussed the production of stop, sbottom,
$\tau$-sneutrino and stau pairs in $e^+e^-$ annihilation in the energy range
$\sqrt{s} = 500$~GeV to $2$~TeV.
We have presented numerical predictions within the Minimal 
Supersymmetric Standard Model for the production cross sections 
and the decay rates and analyzed their SUSY parameter dependence.
If $\tan\beta \,\grts\, 10$, not only the top Yukawa terms,
but also the bottom and tau Yukawa terms have important effects. 
The production 
cross sections as well as the decay rates of stops, sbottoms
and staus depend in a characteristic way on the mixing angles.

A Monte Carlo study of $\eeto\stst$ at $\sqrt{s} = 500$~GeV with 
the decays $\st_{1}\ra c\,\nt_{1}$ and 
$\st_{1}\ra b\,\chp_{1}$ has been performed for $\mst{1} = 180$~GeV,
$\mnt{1} = 100$~GeV, and $\mst{1} = 180$~GeV,
$\mch{1} = 150$~GeV, $\mnt{1} = 60$~GeV, respectively. 
A suitable set of kinematical cuts has been applied to reduce the known
background reactions. In addition, detection regions have been given. 

We have also estimated the experimental accuracies for the masses of 
stops and sbottoms and the stop mixing angle from measurements of the 
polarized cross sections. 
Furthermore, we have made an estimate of the accuracies which can be 
obtained for the soft--breaking SUSY parameters.

In summary, an \ee ~collider --- especially with a polarized $e^{-}$ 
beam --- is an ideal machine for detecting and studying the scalar 
partners of the third generation quarks and leptons.

\section*{Acknowledgements}

We thank our colleagues participating at the Workshop 
``$e^+e^-$ Collisions at TeV Energies: The Physics Potential'' for 
many useful discussions.
This work was supported by the ``Fonds zur F\"orderung der 
wissenschaftlichen Forschung'' of Austria, project no. P10843-PHY.


\clearpage
\section*{Figure Captions}

\setcounter{figure}{1}


\noi
{\bf Fig.~\arabic{figure}a:}~Contour lines for the total 
  cross section of $\eeto\stst$ 
  in f\/b at $\sqrt{s} = 500$~GeV as a function of $\mst{1}$ and 
  $\cos\theta_{\ti t}$. 
  ($\mst{2} = 300$~GeV and $\msg = 300$~GeV.)  \\

\noi
{\bf Fig.~\arabic{figure}b:}~Total cross section of $\eeto\stst$ 
   in f\/b at $\sqrt{s} = 500$~GeV as a function of $\cos\theta_{\ti t}$, 
   for unpolarized (U) as well as left (L) and right (R) polarized 
   $e^{-}$ beams and $\mst{1} = 180$~GeV.
   ($\mst{2} = 300$~GeV and $\msg = 300$~GeV.)  \\

\refstepcounter{figure}   

\noi
{\bf Fig.~\arabic{figure}a:}~Contour lines for the
  total cross section of $\eeto{\st}_2 {\st}_2$ in f\/b at
  $\sqrt{s} = 2$~TeV as a function of $\mst{2}$ and $\cos\theta_{\ti t}$.
  ($\mst{1} = 300$~GeV and $\msg = 700$~GeV.)  \\

\noi
{\bf Fig.~\arabic{figure}b:}~Total cross section of 
  $\eeto {\st}_2 {\st}_2$ in f\/b at $\sqrt{s} = 2$~TeV
  as a function of $\cos\theta_{\ti t}$, for unpolarized (U) as well as
  left (L) and right (R) polarized $e^{-}$ beams and $\mst{2} = 700$~GeV.
  ($\mst{1} = 300$~GeV and $\msg = 700$~GeV.)  \\

\refstepcounter{figure}   

\noi
{\bf Fig.~\arabic{figure}a:}~Contour lines for the
  total cross section of $\eeto\sbsb$ in f\/b at
  $\sqrt{s} = 500$~GeV
  as a function of $\msb{1}$ and $\cos\theta_{\ti b}$.
  ($\msb{2} = 300$~GeV and $\msg = 300$~GeV.)  \\

\noi
{\bf Fig.~\arabic{figure}b:}~Total cross section of $\eeto\sbsb$ in f\/b
   at $\sqrt{s} = 500$~GeV
   as a function of $\cos\theta_{\ti b}$, for unpolarized (U) as well as
   left (L) and right (R) polarized $e^{-}$ beams and $\msb{1} = 180$~GeV.
   ($\msb{2} = 300$~GeV and $\msg = 300$~GeV.)  \\

\refstepcounter{figure}   

\noi
{\bf Fig.~\arabic{figure}a:}~Contour lines for the
  total cross section of $\eeto\staustau$ in f\/b at
  $\sqrt{s} = 500$~GeV
  as a function of $\mstau{1}$ and $\cos\theta_{\ti\tau}$.  \\

\noi
{\bf Fig.~\arabic{figure}b:}~Total cross section of $\eeto\staustau$ in f\/b
   at $\sqrt{s} = 500$~GeV
   as a function of $\cos\theta_{\ti\tau}$, for unpolarized (U) as well as
   left (L) and right (R) polarized $e^{-}$ beams and $\mstau{1} = 180$~GeV.  \\

\refstepcounter{figure}   

\noi
{\bf Fig.~\arabic{figure}:}~Total cross section of 
  $\eeto\ti\nu_\tau\bar{\ti\nu}_\tau$ 
  at $\sqrt{s} = 500$~GeV as a function of $m_{\ti\nu_\tau}$, 
  for unpolarized (U) as well as left (L) and right (R) 
  polarized $e^-$ beams.   \\

\refstepcounter{figure}   

\noi
{\bf Fig.~\arabic{figure}:}~Total cross section of $\eeto\stst$ at 
  $\mst{1} = 180$~GeV
  and $\cos\theta_{\ti t} = 0.7$ as a function of $\sqrt{s}$. 
  ($\mst{2} = 300$~GeV, $\msg = 300$~GeV.)  \\

\refstepcounter{figure}   

\noi
{\bf Fig.~\arabic{figure}:}~Gluon, gluino-top, initial state, and 
  total radiative
  corrections relative to the tree level cross section of $\eeto\stst$ 
  at $\mst{1} = 180$~GeV and $\cos\theta_{\ti t} = 0.7$ as a function 
  of $\sqrt{s}$. ($\mst{2} = 300$~GeV, $\msg = 300$~GeV.)   \\


\refstepcounter{figure}   

\noi
{\bf Fig.~\arabic{figure}a:}~Kinematically allowed parameter domains
in the $M - \mu$ plane for $m_{\tilde t_1} = 180$~GeV and $\tan \beta = 2$
for the decays:
a) $\tilde t_1 \to c \, \tilde \chi^0_1$,
b) $\tilde t_1 \to b \, W^+ \, \tilde \chi^0_1$,
c) $\tilde t_1 \to c \, \tilde \chi^0_2$,
d) $\tilde t_1 \to b \, \tilde \chi^+_1$.
e) $\tilde t_1 \to b \, \tilde \chi^+_2$.
The grey area is covered by LEP2 for $\sqrt{s} =$ 190 GeV.  \\

\noi
{\bf Fig.~\arabic{figure}b:}~Kinematically allowed parameter domains 
in the $M - \mu$ plane for $m_{\tilde b_1} = 180$~GeV and 
$\tan \beta = 30$ for the decays:
a) $\tilde b_1 \to b \, \tilde \chi^0_1$,
b) $\tilde b_1 \to b \, \tilde \chi^0_2$,
c) $\tilde b_1 \to b \, \tilde \chi^0_3$.
The grey area is covered by LEP2 for $\sqrt{s} =$ 190 GeV. \\

\refstepcounter{figure}   

\noi
{\bf Fig.~\arabic{figure}a:}~Branching ratios for the $\tilde t_1$ decays
   as a function of $M$ for $m_{\tilde t_1} = 180$~GeV,
   $\cos \theta_{\tilde t} = 0.7$,
   $\tan \beta = 2$,  and $\mu = -500$~GeV.
   The curves correspond to the following transitions:
   $\circ \hspace{1mm} \tilde t_1 \to c \, \tilde \chi^0_1$,
   \recht $\tilde t_1 \to b \, \tilde \chi^+_-$,
   $\star \hspace{1mm} \tilde t_1 \to b \, W^+ \, \tilde \chi^0_1$.
  The grey area is covered by LEP2 for $\sqrt{s} =$ 190 GeV.  \\

\noi
{\bf Fig.~\arabic{figure}b:}~Branching ratios for the $\tilde b_1$ decays
   as a function of $M$ for $m_{\tilde b_1} = 180$~GeV,
   $\cos \theta_{\tilde b} = 0.7$, $\tan \beta = 30$,
   and $\mu = -500$~GeV.
   The curves correspond to the following transitions:
   $\circ \hspace{1mm} \tilde b_1 \to b \, \tilde \chi^0_1$,
   \rechtl $\tilde b_1 \to b \, \tilde \chi^0_2$.
  The grey area is covered by LEP2 for $\sqrt{s} =$ 190 GeV.  \\

\refstepcounter{figure}   

\noi
{\bf Fig.~\arabic{figure}a:}~Branching ratios for the $\tilde b_1$ decays
   as a function of $M$ for $m_{\tilde b_1} = 180$~GeV,
   $\cos \theta_{\tilde b} = 0.7$,
   $\tan \beta = 30$,  and $\mu = -130$~GeV.
   The curves correspond to the following transitions:
   $\circ \hspace{1mm} \tilde b_1 \to b \, \tilde \chi^0_1$,
   \rechtl $\tilde b_1 \to b \, \tilde \chi^0_2$,
   $\triangle \hspace{1mm} \tilde b_1 \to b \, \tilde \chi^0_3$.
  The grey area is covered by LEP2 for $\sqrt{s} =$ 190 GeV.  \\

\noi
{\bf Fig.~\arabic{figure}b:}~Branching ratios for the $\tilde \tau_1$ decays
   as a function of $M$ for $m_{\tilde \tau_1} = 180$~GeV,
   $\cos \theta_{\tilde \tau} = 0.7$, $\tan \beta = 30$,
   and $\mu = -130$~GeV.
   The curves correspond to the following transitions:
   $\circ \hspace{1mm} \tilde \tau_1 \to \tau \, \tilde \chi^0_1$,
   \rechtl $\tilde \tau_1 \to \tau \, \tilde \chi^0_2$,
   $\triangle \hspace{1mm} \tilde \tau_1 \to \tau \, \tilde \chi^0_3$,
   $\Diamond \hspace{1mm} \tilde \tau_1 \to \tau \, \tilde \chi^0_4$,
   \recht $\tilde \tau_1 \to \nu_{\tau} \, \tilde \chi^-_1$.
  The grey area is covered by LEP2 for $\sqrt{s} =$ 190 GeV.  \\

\noi
{\bf Fig.~\arabic{figure}c:}~Branching ratios for the $\tilde \nu_{\tau}$
   decays as a function of $M$ for $m_{\tilde \nu_{\tau}} = 180$~GeV,
   $\tan \beta = 30$,  and $\mu = -130$~GeV.
   The curves correspond to the following transitions:
   $\circ \hspace{1mm} \tilde \nu_{\tau}
                       \to \nu_{\tau} \, \tilde \chi^0_1$,
   \rechtl $\tilde \nu_{\tau} \to \nu_{\tau} \, \tilde \chi^0_2$,
   $\triangle \hspace{1mm} \tilde \nu_{\tau}
                       \to \nu_{\tau} \, \tilde \chi^0_3$,
   $\Diamond \hspace{1mm} \tilde \nu_{\tau}
                       \to \nu_{\tau} \, \tilde \chi^0_4$,
   \recht $\tilde \nu_{\tau} \to \tau \, \tilde \chi^+_1$.
  The grey area is covered by LEP2 for $\sqrt{s} =$ 190 GeV.  \\

\refstepcounter{figure}   

\noi
{\bf Fig.~\arabic{figure}:}~Branching ratio for the decays of
the heavier stop as
a function of $M$ for $\tan \beta= 2$ and $\mu = -1$~TeV. \\
(a)~shows the decays into fermions:
   --- \hspace{1mm} $ \tilde t_2 \to t \, \tilde g$,
   $\circ \hspace{1mm} \tilde t_2 \to t \, \tilde \chi^0_1$,
   \rechtl $\tilde t_2 \to t \, \tilde \chi^0_2$,
   \recht $\tilde t_2 \to b \, \tilde \chi^+_1$.  \\
(b)~shows the decays into bosons:
   $\circ \hspace{1mm} \tilde t_2 \to Z \, \tilde t_1$,
   \rechtl $\tilde t_2 \to h^0 \, \tilde t_1$,
   $\triangle \hspace{1mm} \tilde t_2 \to H^0 \, \tilde t_1$, 
   $\Diamond \hspace{1mm} \tilde t_2 \to A^0 \, \tilde t_1$,
   \recht $\tilde t_2 \to W^+ \, \tilde b_1$.
The grey area is covered by LEP2 for $\sqrt{s} =$ 190 GeV.  \\

\refstepcounter{figure}   

\noi
{\bf Fig.~\arabic{figure}:}~Branching ratio for the decays of
the heavier sbottom as
a function of $M$ for $\tan \beta= 30$ and $\mu = -1$~TeV. \\
(a)~shows the decays into fermions:
   --- \hspace{1mm} $ \tilde b_2 \to b \, \tilde g$,
   $\circ \hspace{1mm} \tilde b_2 \to b \, \tilde \chi^0_1$,
   \rechtl $\tilde b_2 \to b \, \tilde \chi^0_2$,
   \recht $\tilde b_2 \to t \, \tilde \chi^-_1$.  \\
(b)~shows the decays into bosons:
   $\circ \hspace{1mm} \tilde b_2 \to Z \, \tilde b_1$,
   \rechtl $\tilde b_2 \to h^0 \, \tilde b_1$,
   $\triangle \hspace{1mm} \tilde b_2 \to H^0 \, \tilde b_1$,  
   $\Diamond \hspace{1mm} \tilde b_2 \to A^0 \, \tilde b_1$,
   \recht $\tilde b_2 \to W^- \, \tilde t_1$,
   $\bullet \hspace{1mm} \tilde b_2 \to H^- \, \tilde t_1$.
The grey area is covered by LEP2 for $\sqrt{s} =$ 190 GeV.  \\

\refstepcounter{figure}   

\noi
{\bf Fig.~\arabic{figure}:}~Branching ratio for the decays of
the heavier stau as
a function of $M$ for $\tan \beta= 30$ and $\mu = -1$~TeV.  \\
(a) shows the decays into fermions:
   $\circ \hspace{1mm} \tilde \tau_2 \to \tau \, \tilde \chi^0_1$,
   \rechtl $\tilde \tau_2 \to \tau \, \tilde \chi^0_2$,
   \recht $\tilde \tau_2 \to \nu_{\tau} \, \tilde \chi^-_1$.  \\
(b) shows the decays into bosons:
   $\circ \hspace{1mm} \tilde \tau_2 \to Z \, \tilde \tau_1$,
   \rechtl $\tilde \tau_2 \to h^0 \, \tilde \tau_1$,
   $\triangle \hspace{1mm} \tilde \tau_2 \to H^0 \, \tilde \tau_1$,
   $\Diamond \hspace{1mm} \tilde \tau_2 \to A^0 \, \tilde \tau_1$.
The grey area is covered by LEP2 for $\sqrt{s} =$ 190 GeV.  \\

\refstepcounter{figure}   

\noi
{\bf Fig.~\arabic{figure}:}~Branching ratio for the decays of
the tau sneutrino as
a function of $M$ for $\tan \beta= 30$ and $\mu = -1$~TeV.  \\
(a)~shows the decays into fermions:
   $\circ \hspace{1mm} \tilde \nu_{\tau} \to \nu_{\tau} \, \tilde \chi^0_1$,
   \rechtl $\tilde \nu_{\tau} \to \nu_{\tau} \, \tilde \chi^0_2$,
   \recht $\tilde \nu_{\tau} \to \tau \, \tilde \chi^+_1$.  \\
(b)~shows the decays into bosons:
   \recht $\tilde \nu_{\tau} \to W^+ \, \tilde \tau_1$,
   $\bullet \hspace{1mm} \tilde \nu_{\tau} \to H^+ \, \tilde \tau_1$.
The grey area is covered by LEP2 for $\sqrt{s} =$ 190 GeV.  \\


\refstepcounter{figure}   

\noi
{\bf Fig.~\arabic{figure}:}~Background reactions and their cross sections for
        $\sqrt{s}=\gev{500}$.  \\
\refstepcounter{figure}

\noi
{\bf Fig.~\arabic{figure}:}~$E_{vis}/\sqrt{s}<0.4$ for 
         \chichi, qq, WW, eW$\nu$, tt, ZZ, eeZ.  \\
\refstepcounter{figure}

\noi
{\bf Fig.~\arabic{figure}:}~$m_{\mathrm{inv}}>\gev{120}$
                for \chichi, qq, WW, eW$\nu$, tt, ZZ, eeZ.  \\
\refstepcounter{figure}

\noi
{\bf Fig.~\arabic{figure}:}~Sensitivity for an 
$\eeto\st_1\bar{\st}_1 \rightarrow \nt_1 c\nt_1\bar{c}$ 
signal.
Open histograms show the simulated signal, solid
and hatched histograms show the remaining background
after all selection cuts are applied. \\
\refstepcounter{figure}

\noi
{\bf Fig.~\arabic{figure}:}~Thrust $<0.85$ 
         for \chipchip, qq, WW, eW$\nu$, tt, ZZ, eeZ.  \\
\refstepcounter{figure}

\noi
{\bf Fig.~\arabic{figure}:}~Number of cluster $\geq 60$
         for \chipchip, qq, WW, eW$\nu$, tt, ZZ, eeZ.  \\

\refstepcounter{figure}   

\noi
{\bf Fig.~\arabic{figure}:}~Sensitivity for an
$\eeto\st_1\bar{\st}_1 \rightarrow \chp_1 b\chm_1\bar{b}$ 
signal. Open histograms show the simulated signal, solid
and hatched histograms show the remaining background
after all selection cuts are applied. \\

\refstepcounter{figure}   

\noi
{\bf Fig.~\arabic{figure}:}~Detection confidence levels.
(a) \chipchip\ channel. (b) \chichi\ channel.  \\

\refstepcounter{figure}   

\noi
{\bf Fig.~\arabic{figure}:}~~Discovery region for $\sqrt{s} = 500$~GeV 
and ${\cal L} = 50~\fb^{-1}$. In the shaded area a $3\sigma$ effect 
is expected. \\

\refstepcounter{figure}   

\noi
{\bf Fig.~\arabic{figure}:}~Discovery region for $\sqrt{s} = 800$~GeV 
and ${\cal L} = 200~\fb^{-1}$. In the shaded area a $3\sigma$ effect 
is expected.\\


\refstepcounter{figure}   

\noi
{\bf Fig.~\arabic{figure}:}~Error bands for the total tree--level cross
  section of $e^+ e^- \to \st_1 \st_1$ in f\/b at
  $\sqrt{s} = 400$~GeV and $\sqrt{s} = 500$~GeV
  as a function of $m_{\st_1}$ and $\cos\theta_{\st}$.
  The dot corresponds to $m_{\st_1} = 180$~GeV and  
  $\cos\theta_{\st} = 0.57$. 
  The error bands are defined by 
  $(\sigma_{400}, \Delta \sigma_{400}) = (18.2, 4.1)$~f\/b and
  $(\sigma_{500}, \Delta \sigma_{500}) = (47.4, 5.5)$~f\/b.  \\

\refstepcounter{figure}   

\noi
{\bf Fig.~\arabic{figure}:}~Error bands (dashed) and the corresponding 
  error ellipse as a function of $m_{\st_1}$ and $\cos\theta_{\st}$ 
  for the total tree-level cross sections of $e^+ e^- \to \st_1 \st_1$ 
  in f\/b at $\sqrt{s} = 500$~GeV with 90\% left-- and right--polarized
  electron beam.
  The dot corresponds to $m_{\st_1} = 180$~GeV and 
  $\cos\theta_{\st} = 0.57$. 
  The error bands are defined by 
  $(\sigma_L, \Delta \sigma_L) = (48.6, 6.0)$~f\/b and
  $(\sigma_R, \Delta \sigma_R) = (46.1, 4.9)$~f\/b.  \\

\refstepcounter{figure}   


\clearpage 

\setcounter{figure}{1}

\noi
\begin{minipage}[t]{73mm}   
{\setlength{\unitlength}{1mm}
\begin{picture}(73,76)
\put(3,4){\mbox{\psfig{figure=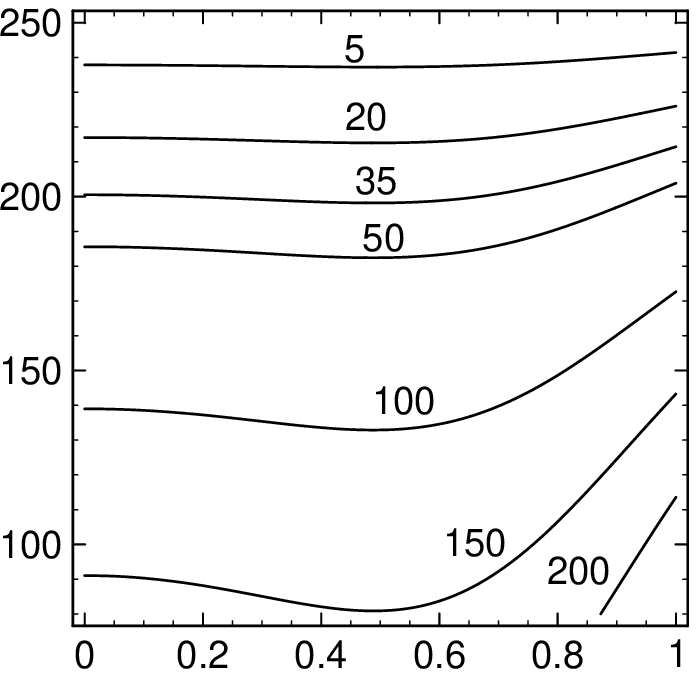,height=6.6cm}}}
\put(3,70){\makebox(0,0)[bl]{{$\mst{1}$~[GeV]}}}
\put(69.5, 0.3){\makebox(0,0)[br]{{$\cos\theta_{\ti t}$}}}
\end{picture}}
\bce{\large{\bf Fig.~\arabic{figure}a}}\ece
\end{minipage}
  \hspace{3mm}
\begin{minipage}[t]{73mm}   
{\setlength{\unitlength}{1mm}
\begin{picture}(73,76)
\put(3,4){\mbox{\psfig{figure=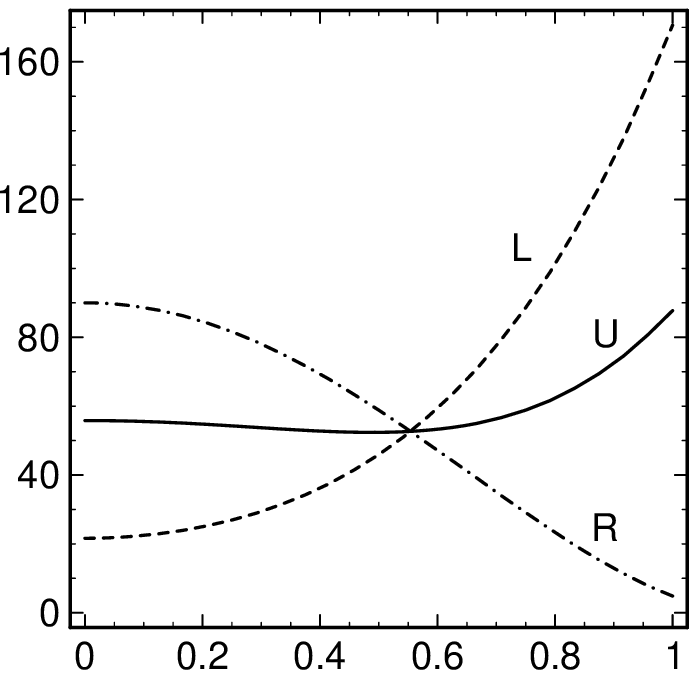,height=6.6cm}}}
\put(69.5,0.5){\makebox(0,0)[br]{{$\cos\theta_{\ti t}$}}}
\put(3,70){\makebox(0,0)[bl]{{$\sigma(\stst )$~[f\/b]}}}
\end{picture}}
\bce{\large{\bf Fig.~\arabic{figure}b}}\ece
\end{minipage}
\refstepcounter{figure}     
\vspace{15mm}

\noi
\begin{minipage}[t]{73mm}   
{\setlength{\unitlength}{1mm}
\begin{picture}(73,76)
\put(3,4){\mbox{\psfig{figure=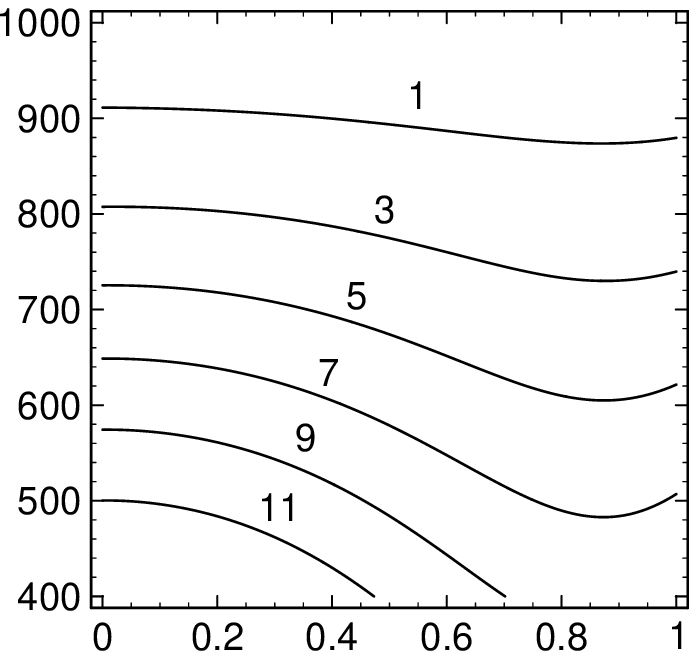,height=6.6cm}}}
\put(3,70){\makebox(0,0)[bl]{$\mst{2}$~[GeV]}}
\put(69.5, 0.3){\makebox(0,0)[br]{$\cos\theta_{\ti t}$}}
\end{picture}}
\bce{\large{\bf Fig.~\arabic{figure}a}}\ece
\end{minipage}
   \hspace{3mm}
\begin{minipage}[t]{73mm}   
{\setlength{\unitlength}{1mm}
\begin{picture}(73,76)
\put(5,4){\mbox{\psfig{figure=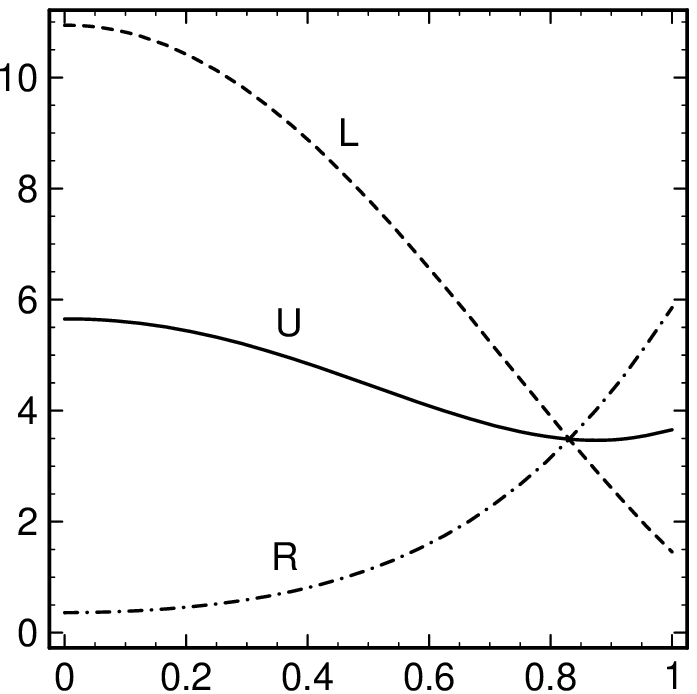,height=6.4cm}}}
\put(69.5,-0.5){\makebox(0,0)[br]{$\cos\theta_{\ti t}$}}
\put(3,70){\makebox(0,0)[bl]{$\sigma({\st}_2 {\st}_2)$~[f\/b]}}
\end{picture}}
\bce{\large{\bf Fig.~\arabic{figure}b}}\ece
\end{minipage}
\refstepcounter{figure}

\clearpage 

\noi
\begin{minipage}[t]{73mm}   
{\setlength{\unitlength}{1mm}
\begin{picture}(73,76)
\put(3,4){\mbox{\psfig{figure=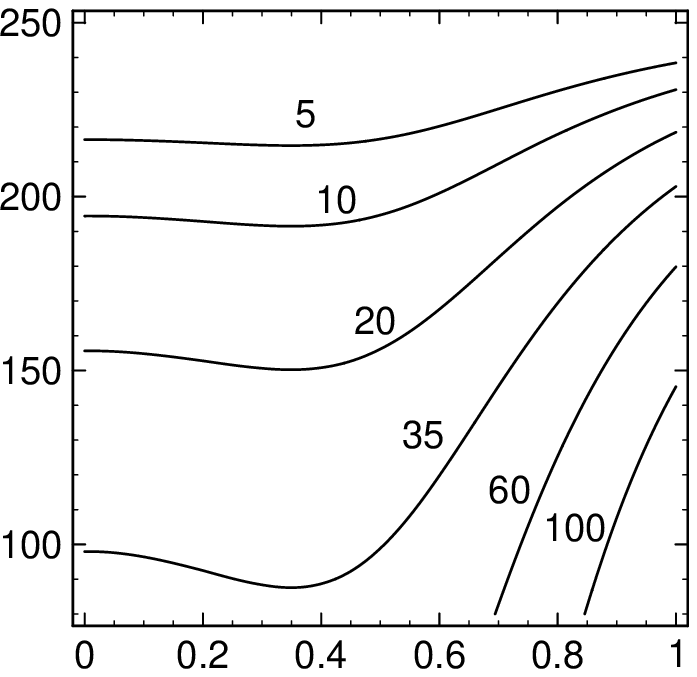,height=6.6cm}}}
\put(3,70){\makebox(0,0)[bl]{$\msb{1}$~[GeV]}}
\put(69.5, 0.3){\makebox(0,0)[br]{$\cos\theta_{\ti b}$}}
\end{picture}}
\bce{\large{\bf Fig.~\arabic{figure}a}}\ece
\end{minipage}
   \hspace{3mm}
\begin{minipage}[t]{73mm}   
{\setlength{\unitlength}{1mm}
\begin{picture}(73,76)
\put(3,4){\mbox{\psfig{figure=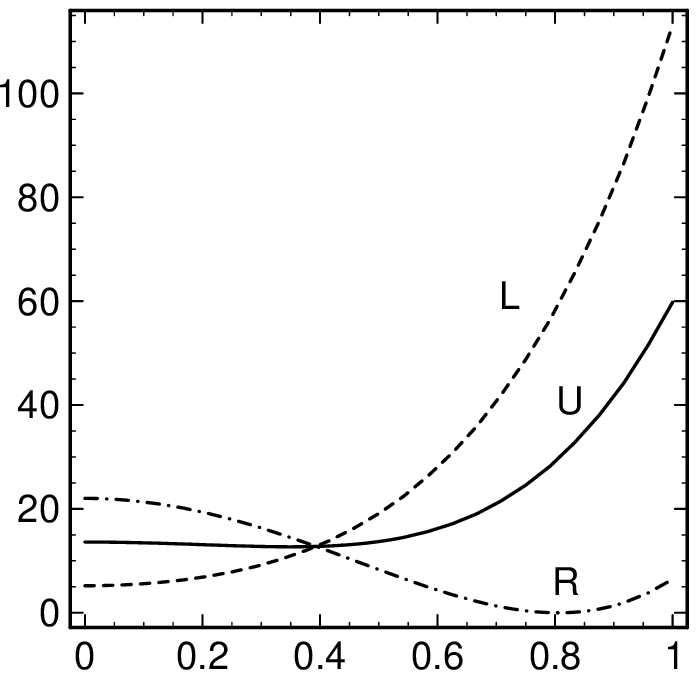,height=6.6cm}}}
\put(69.5,0.5){\makebox(0,0)[br]{$\cos\theta_{\ti b}$}}
\put(3,70){\makebox(0,0)[bl]{$\sigma(\sbsb )$~[f\/b]}}
\end{picture}}
\bce{\large{\bf Fig.~\arabic{figure}b}}\ece
\end{minipage}
\refstepcounter{figure}
\vspace{15mm}

\noi
\begin{minipage}[t]{73mm}   
{\setlength{\unitlength}{1mm}
\begin{picture}(73,76)
\put(3,4){\mbox{\psfig{figure=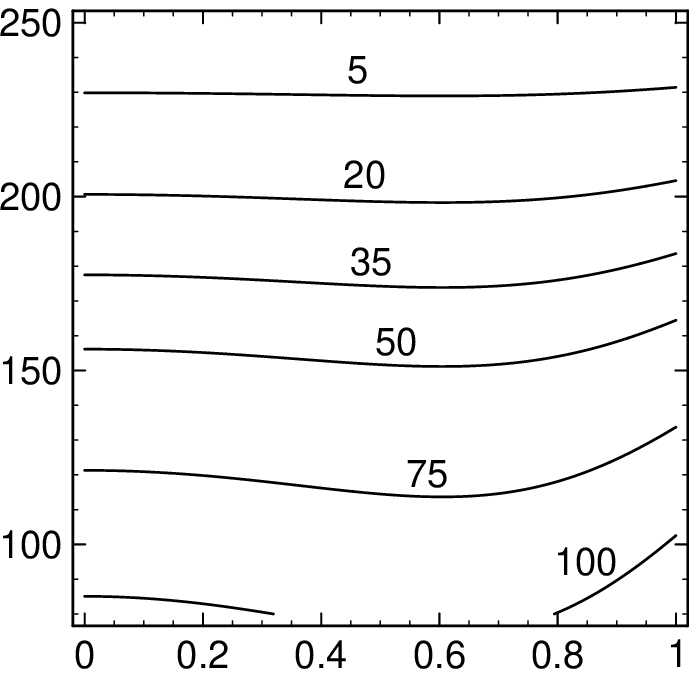,height=6.6cm}}}
\put(3,70){\makebox(0,0)[bl]{$\mstau{1}$~[GeV]}}
\put(69.5, 0.3){\makebox(0,0)[br]{$\cos\theta_{\ti\tau}$}}
\end{picture}}
\bce{\large{\bf Fig.~\arabic{figure}a}}\ece
\end{minipage}
   \hspace{3mm}
\begin{minipage}[t]{73mm}   
{\setlength{\unitlength}{1mm}
\begin{picture}(73,76)
\put(5,4){\mbox{\psfig{figure=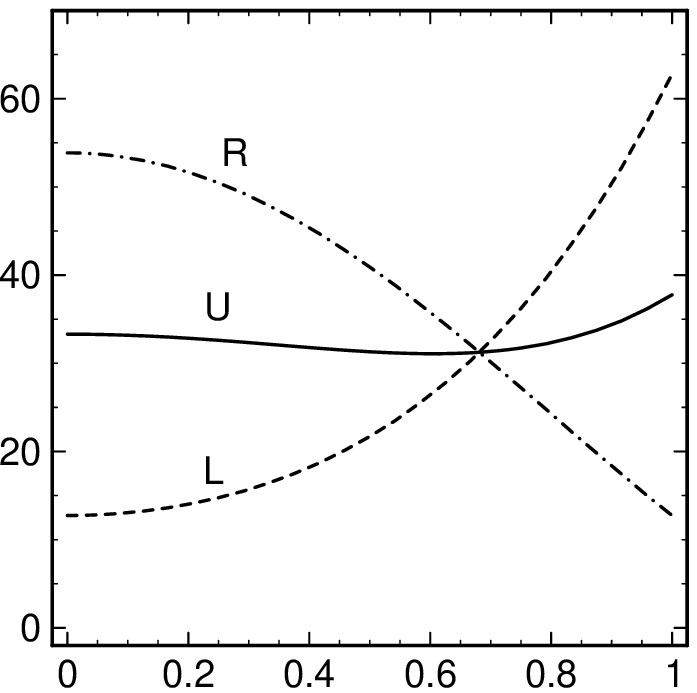,height=6.4cm}}}
\put(69.5,0.5){\makebox(0,0)[br]{$\cos\theta_{\ti\tau}$}}
\put(3,70){\makebox(0,0)[bl]{$\sigma(\staustau )$~[f\/b]}}
\end{picture}}
\bce{\large{\bf Fig.~\arabic{figure}b}}\ece
\end{minipage}
\refstepcounter{figure}

\clearpage 

\noi
\begin{center}
\begin{minipage}[t]{73mm}   
{\setlength{\unitlength}{1mm}
\begin{picture}(73,76)    
\put(3,4){\mbox{\psfig{figure=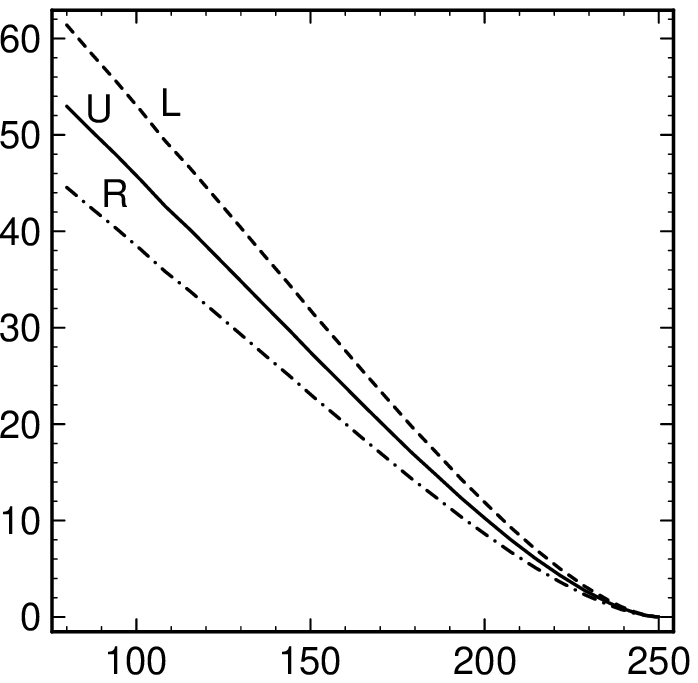,height=6.6cm}}}
\put(69.5,-0.5){\makebox(0,0)[br]{$m_{\ti\nu_\ell}$~[GeV]}}
\put(3,71){\makebox(0,0)[bl]{$\sigma(\ti\nu_\ell\bar{\ti \nu}_\ell)$~[f\/b]}}
\end{picture}}
\bce{\large{\bf Fig.~\arabic{figure}}}\ece
\end{minipage}
\refstepcounter{figure}
\end{center}
\vspace{15mm}

\noi
\begin{minipage}[t]{73mm}   
{\setlength{\unitlength}{1mm}
\begin{picture}(73,76)
\put(3,4){\mbox{\psfig{figure=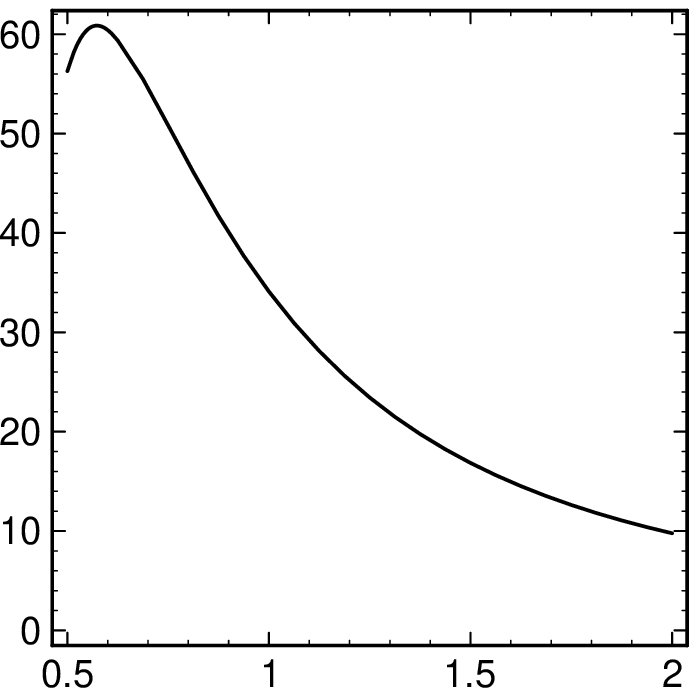,height=6.6cm}}}
\put(69.5,0.){\makebox(0,0)[br]{$\sqrt{s}$~[TeV]}}
\put(3,71){\makebox(0,0)[bl]{$\sigma(\stst )$~[f\/b]}}
\end{picture}}
\bce{\large{\bf Fig.~\arabic{figure}}}\ece
\end{minipage}
\refstepcounter{figure}
  \hspace{3mm}
\begin{minipage}[t]{73mm}   
{\setlength{\unitlength}{1mm}
\begin{picture}(73,76)
\put(3,4){\mbox{\psfig{figure=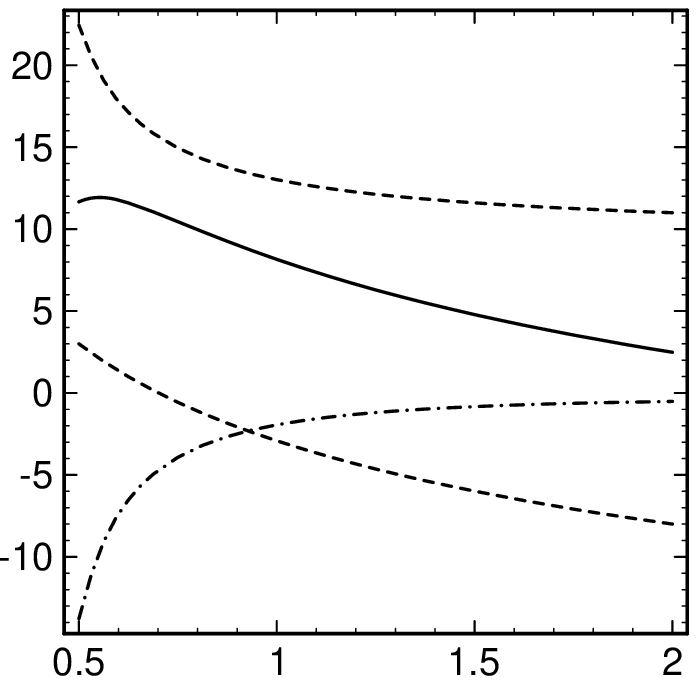,height=6.6cm}}}
\put(69.5,0.){\makebox(0,0)[br]{$\sqrt{s}$~[TeV]}}
\put(3,71){\makebox(0,0)[bl]{$\Delta\sigma/\sigma^{\mbox{tree
}}$~[\%]}}
\put(36,52.5){\makebox(0,0)[bl]{\small gluon}}
\put(43,42){\makebox(0,0)[bl]{\small total}}
\put(50, 31.5){\makebox(0,0)[br]{\small gluino}}
\put(53,23){\makebox(0,0)[bl]{\small ISR}}
\end{picture}}
\bce{\large{\bf Fig.~\arabic{figure}}}\ece
\end{minipage}
\refstepcounter{figure}

\clearpage 

\noi
\begin{minipage}[t]{73mm}   
{\setlength{\unitlength}{1mm}
\begin{picture}(73,76)                        
\put(3,4){\mbox{\epsfig{figure=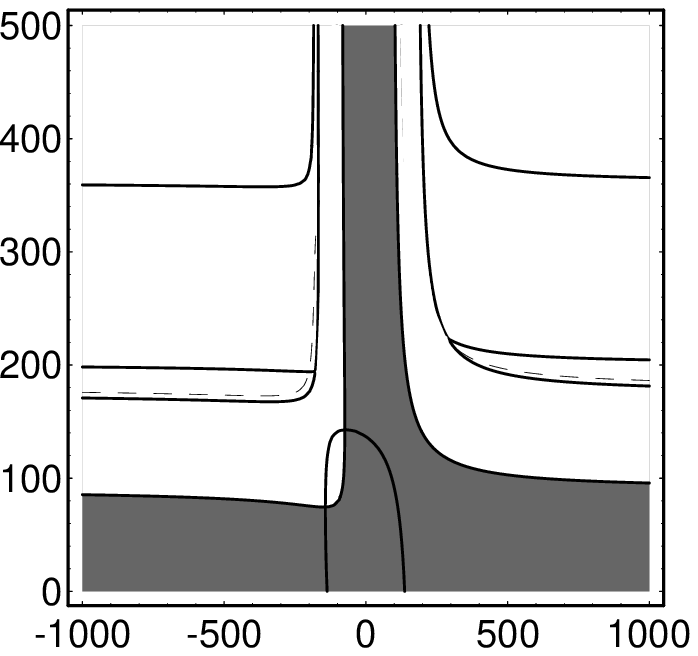,height=6.6cm}}}
\put(69.5,1){\makebox(0,0)[br]{$\mu$~[GeV]}}
\put(3,70){\makebox(0,0)[bl]{$M$~[GeV]}}
\put(12,58.5){{\small $m_{\tilde t_1} < m_{\tilde \chi^0_1}$}}
\put(47,59.5){{\small $m_{\tilde t_1} < m_{\tilde \chi^0_1}$}}
\put(20,43){{\small a}}
\put(60.5,43){{\small a}}
\put(20,34){{\small b}}
\put(22.2,34.9){\vector(0,-1){4}}
\put(50.5,35.8){{\small b}}
\put(53,37.0){\vector(0,-1){4}}
\put(30,34){{\small c}}
\put(29.5,34.1){\vector(0,-1){4}}
\put(60.5,35){{\small c}}
\put(60.0,35.5){\vector(0,-1){4}}
\put(20,25){{\small d}}
\put(60.5,25){{\small d}}
\put(34.5,23){{\small e}}
\put(34,15){{\small LEP2 }}
\end{picture}}
\bce{\large{\bf Fig.~\arabic{figure}a}}\ece
\end{minipage}
  \hspace{3mm}
\begin{minipage}[t]{73mm}   
{\setlength{\unitlength}{1mm}
\begin{picture}(73,76)                        
\put(3,4){\mbox{\epsfig{figure=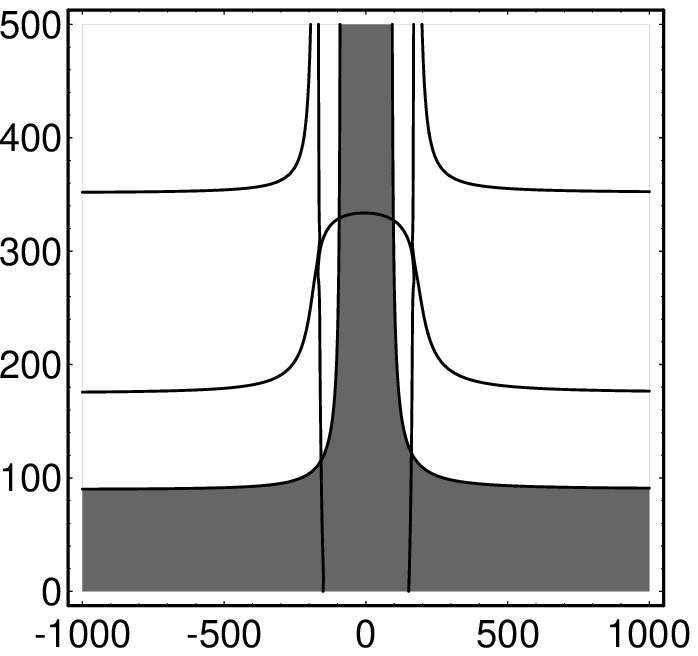,height=6.6cm}}}
\put(69.5,1){\makebox(0,0)[br]{$\mu$~[GeV]}}
\put(3,70){\makebox(0,0)[bl]{$M$~[GeV]}}
\put(12,58.5){{\small $m_{\tilde b_1} < m_{\tilde \chi^0_1}$}}
\put(47,59.5){{\small $m_{\tilde b_1} < m_{\tilde \chi^0_1}$}}
\put(20,39.5){{\small a}}
\put(55,39.5){{\small a}}
\put(20,24.5){{\small b}}
\put(55,24.5){{\small b}}
\put(33.8,39.5){{\small c}}
\put(41,39.5){{\small c}}
\put(34,15){{\small LEP2 }}
\end{picture}}
\bce{\large{\bf Fig.~\arabic{figure}b}}\ece
\end{minipage}
\vspace{15mm}
\refstepcounter{figure}

\noi
\begin{minipage}[t]{73mm}   
{\setlength{\unitlength}{1mm}
\begin{picture}(73,76)                        
\put(3,4){\mbox{\epsfig{figure=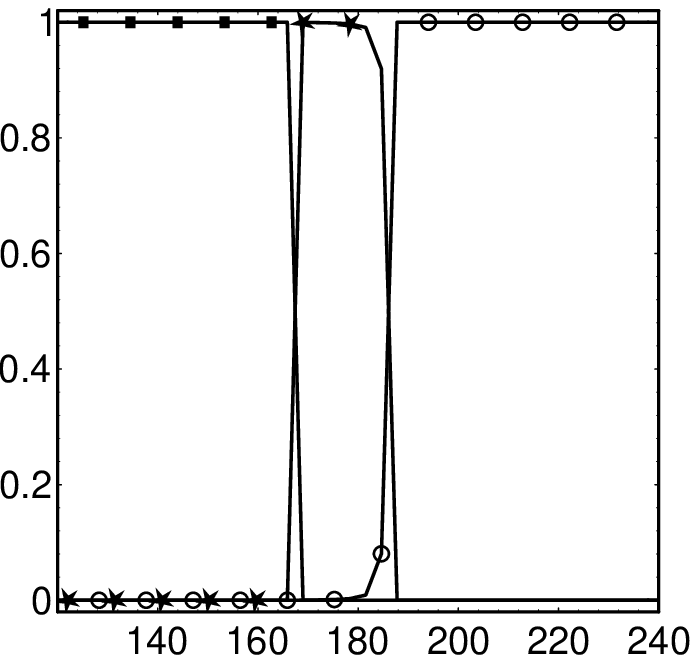,height=6.6cm}}}
\put(69.5,1){\makebox(0,0)[br]{$M$~[GeV]}}
\put(3,70){\makebox(0,0)[bl]{BR $(\tilde t_{1})$~[\%]}}
\end{picture}}
\bce{\large{\bf Fig.~\arabic{figure}a}}\ece
\end{minipage}
\hspace{3mm}
\begin{minipage}[t]{73mm}   
{\setlength{\unitlength}{1mm}
\begin{picture}(73,76)
\put(3,4){\mbox{\epsfig{figure=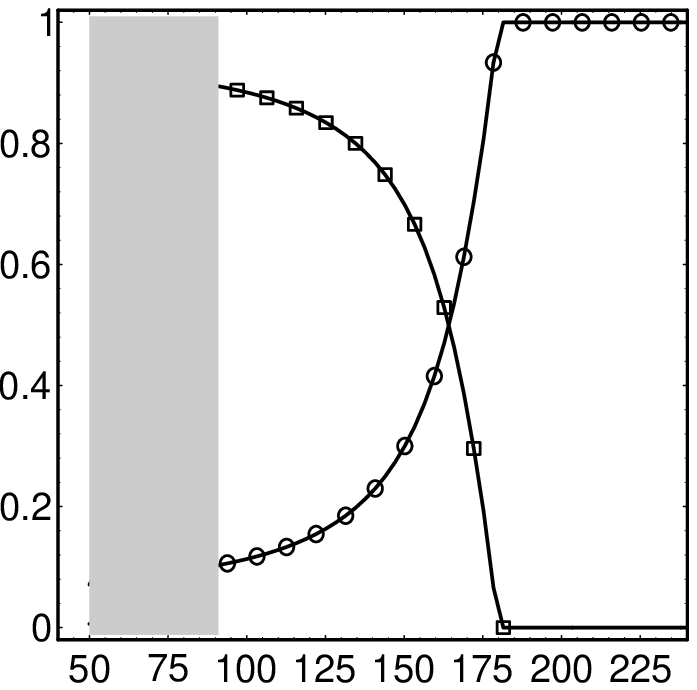,height=6.6cm}}}
\put(69.5,0){\makebox(0,0)[br]{$M$~[GeV]}}
\put(3,70){\makebox(0,0)[bl]{BR $(\tilde b_{1})$~[\%]}}
\end{picture}}
\bce{\large{\bf Fig.~\arabic{figure}b}}\ece
\end{minipage}
\vspace{5mm}
\refstepcounter{figure}

\noi
\begin{minipage}[t]{73mm}   
{\setlength{\unitlength}{1mm}
\begin{picture}(73,76)                        
\put(3,4){\mbox{\epsfig{figure=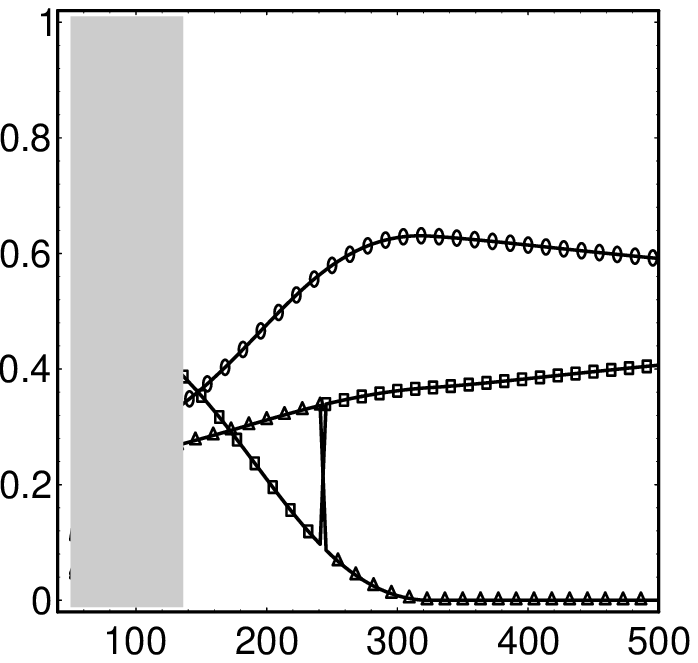,height=6.6cm}}}
\put(69.5,1){\makebox(0,0)[br]{$M$~[GeV]}}
\put(3,70){\makebox(0,0)[bl]{BR $(\tilde b_{1})$~[\%]}}
\end{picture}}
\bce{\large{\bf Fig.~\arabic{figure}a}}\ece
\end{minipage}
  \hspace{3mm}
\begin{minipage}[t]{73mm}   
{\setlength{\unitlength}{1mm}
\begin{picture}(73,76)
\put(3,4){\mbox{\epsfig{figure=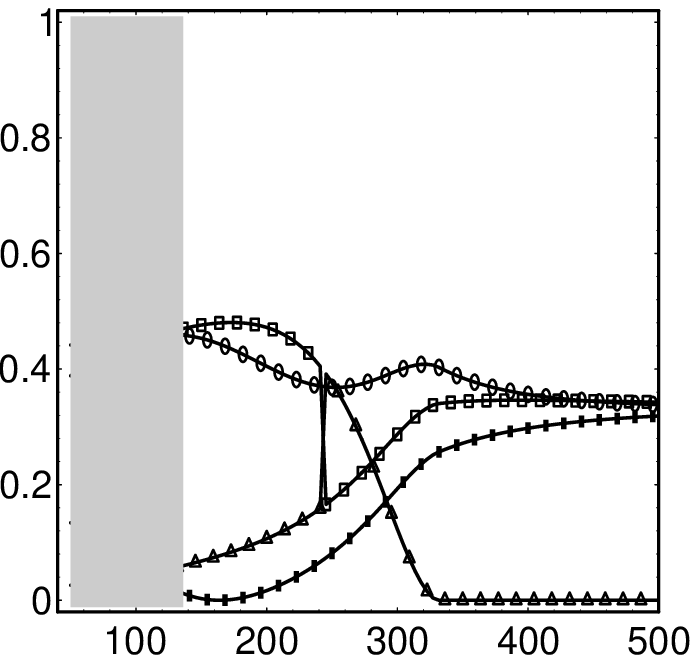,height=6.6cm}}}
\put(69.5,1){\makebox(0,0)[br]{$M$~[GeV]}}
\put(3,70){\makebox(0,0)[bl]{BR $(\tilde \tau_{1})$~[\%]}}
\end{picture}}
\bce{\large{\bf Fig.~\arabic{figure}b}}\ece
\end{minipage}
\vspace{15mm}

\noi
\begin{center}
\begin{minipage}[t]{73mm}   
{\setlength{\unitlength}{1mm}
\begin{picture}(73,76)
\put(3,4){\mbox{\epsfig{figure=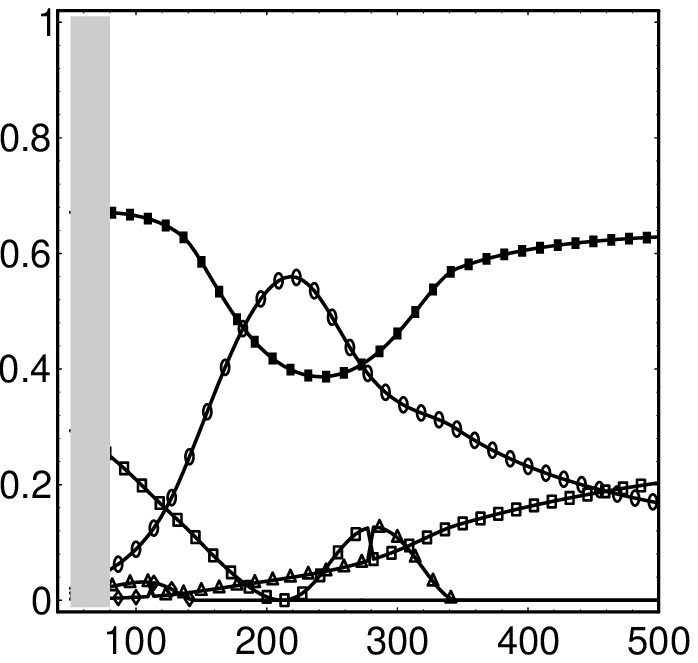,height=6.6cm}}}
\put(69.5,1){\makebox(0,0)[br]{$M$~[GeV]}}
\put(3,70){\makebox(0,0)[bl]{BR $(\tilde \nu_{\tau})$~[\%]}}
\end{picture}}
\bce{\large{\bf Fig.~\arabic{figure}c}}\ece
\end{minipage}
\end{center}
\refstepcounter{figure}
\vspace{5mm}

\noindent
\begin{minipage}[t]{150mm}   
{\setlength{\unitlength}{1mm}
\begin{picture}(150,76)
\put(3,4){\mbox{\epsfig{figure=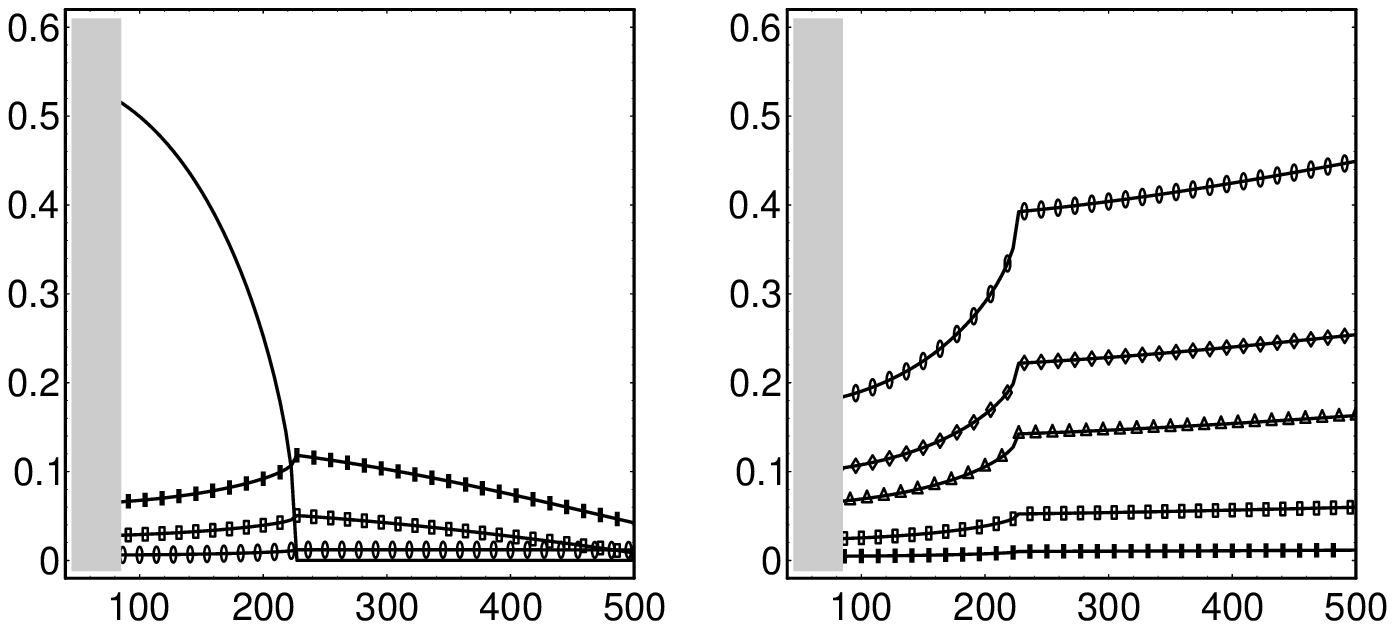,height=6.6cm}}}
\put(0,71){a)}
\put(69.5,3){\makebox(0,0)[br]{$M$~[GeV]}}
\put(8,70){\makebox(0,0)[bl]{$BR(\tilde t_2)$~[\%]}}
\put(70,71){b)}
\put(138.5,3){\makebox(0,0)[br]{$M$~[GeV]}}
\put(77,70){\makebox(0,0)[bl]{$BR(\tilde t_2)$~[\%]}}
\end{picture}}
\bce{\large{\bf Fig.~\arabic{figure}}}\ece
\end{minipage}
\refstepcounter{figure}
\vspace{15mm}

\noindent
\begin{minipage}[t]{150mm}   
{\setlength{\unitlength}{1mm}
\begin{picture}(150,76)
\put(3,4){\mbox{\epsfig{figure=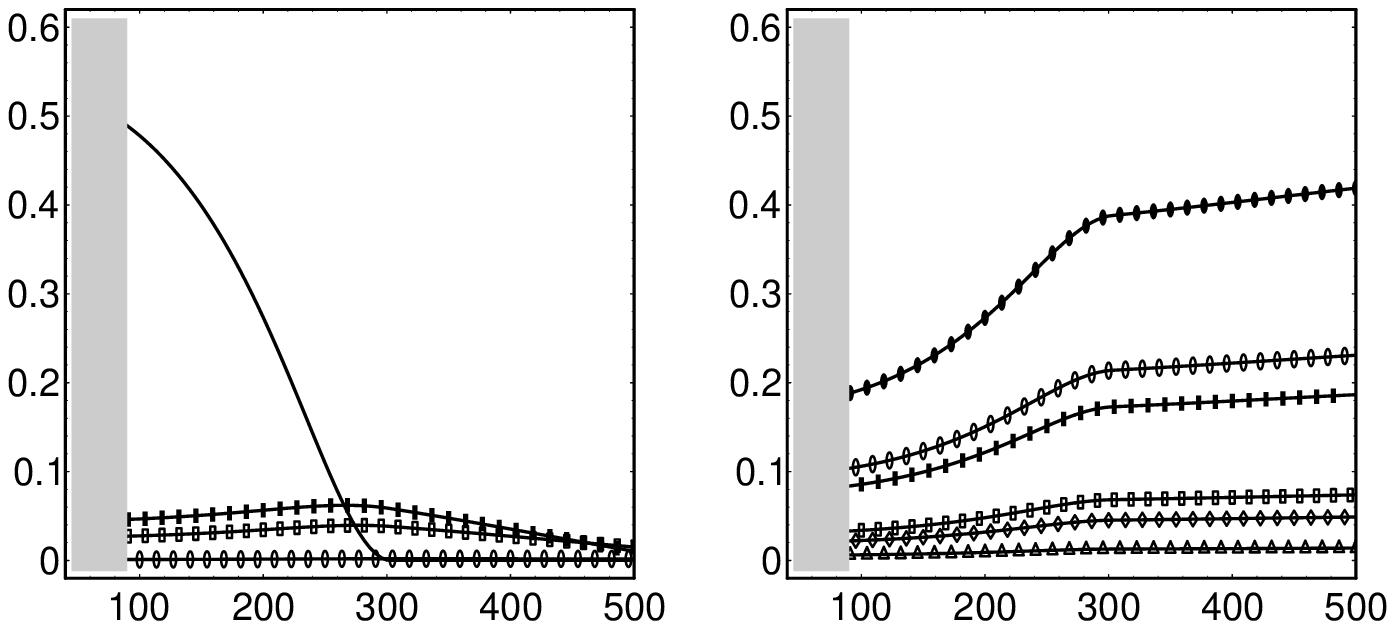,height=6.6cm}}}
\put(0,71){a)}
\put(69.5,3){\makebox(0,0)[br]{$M$~[GeV]}}
\put(8,70){\makebox(0,0)[bl]{$BR(\tilde b_2)$~[\%]}}
\put(70,71){b)}
\put(138.5,3){\makebox(0,0)[br]{$M$~[GeV]}}
\put(77,70){\makebox(0,0)[bl]{$BR(\tilde b_2)$~[\%]}}
\end{picture}}
\bce{\large{\bf Fig.~\arabic{figure}}}\ece
\end{minipage}
\refstepcounter{figure}

\noindent
\begin{minipage}[t]{150mm}   
{\setlength{\unitlength}{1mm}
\begin{picture}(150,76)
\put(3,4){\mbox{\epsfig{figure=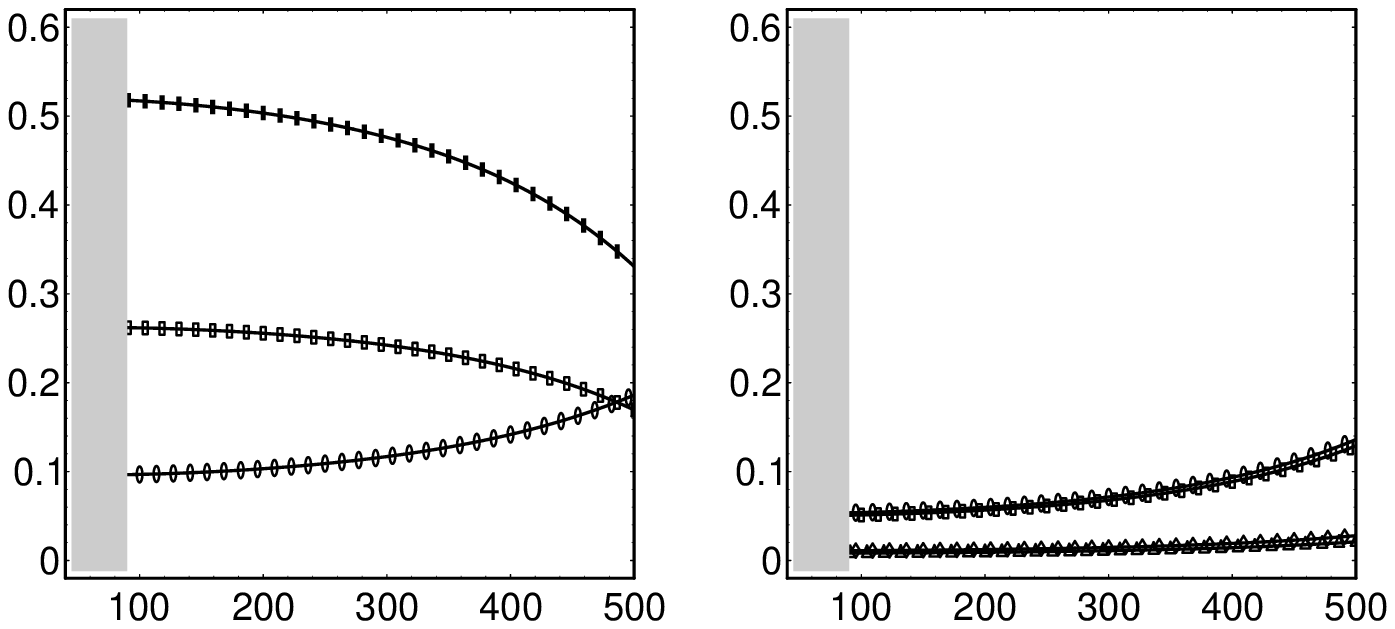,height=6.6cm}}}
\put(0,71){a)}
\put(69.5,3){\makebox(0,0)[br]{$M$~[GeV]}}
\put(8,70){\makebox(0,0)[bl]{$BR(\tilde \tau_2)$~[\%]}}
\put(70,71){b)}
\put(138.5,3){\makebox(0,0)[br]{$M$~[GeV]}}
\put(77,70){\makebox(0,0)[bl]{$BR(\tilde \tau_2)$~[\%]}}
\end{picture}}
\bce{\large{\bf Fig.~\arabic{figure}}}\ece
\end{minipage}
\refstepcounter{figure}
\vspace{15mm}
 
\noindent
\begin{minipage}[t]{150mm}   
{\setlength{\unitlength}{1mm}
\begin{picture}(150,76)
\put(3,4){\mbox{\epsfig{figure=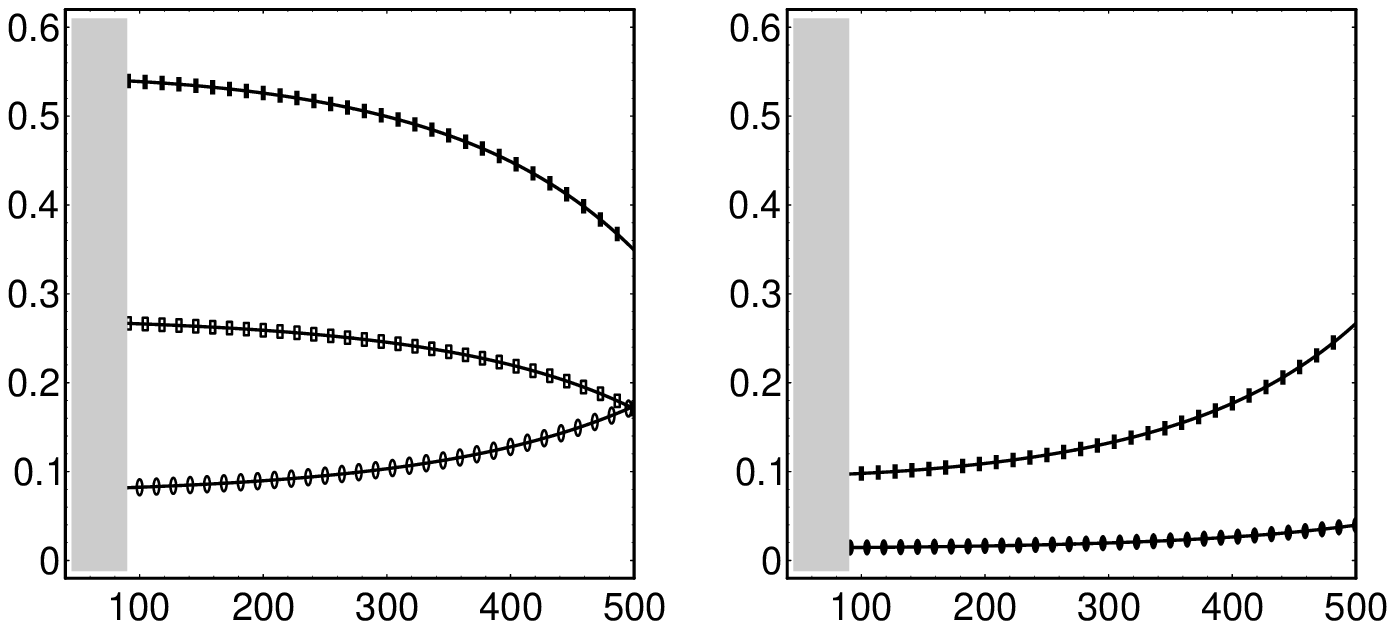,height=6.6cm}}}
\put(0,71){a)}
\put(69.5,3){\makebox(0,0)[br]{$M$~[GeV]}}
\put(8,70){\makebox(0,0)[bl]{$BR(\tilde \nu_{\tau})$~[\%]}}
\put(70,71){b)}
\put(138.5,3){\makebox(0,0)[br]{$M$~[GeV]}}
\put(77,70){\makebox(0,0)[bl]{$BR(\tilde \nu_{\tau})$~[\%]}}
\end{picture}}
\bce{\large{\bf Fig.~\arabic{figure}}}\ece
\end{minipage}
\refstepcounter{figure}

\clearpage 

\noindent
{\setlength{\unitlength}{1mm}    
\begin{picture}(150,100)
\put(3,4){\mbox{\epsfig{figure=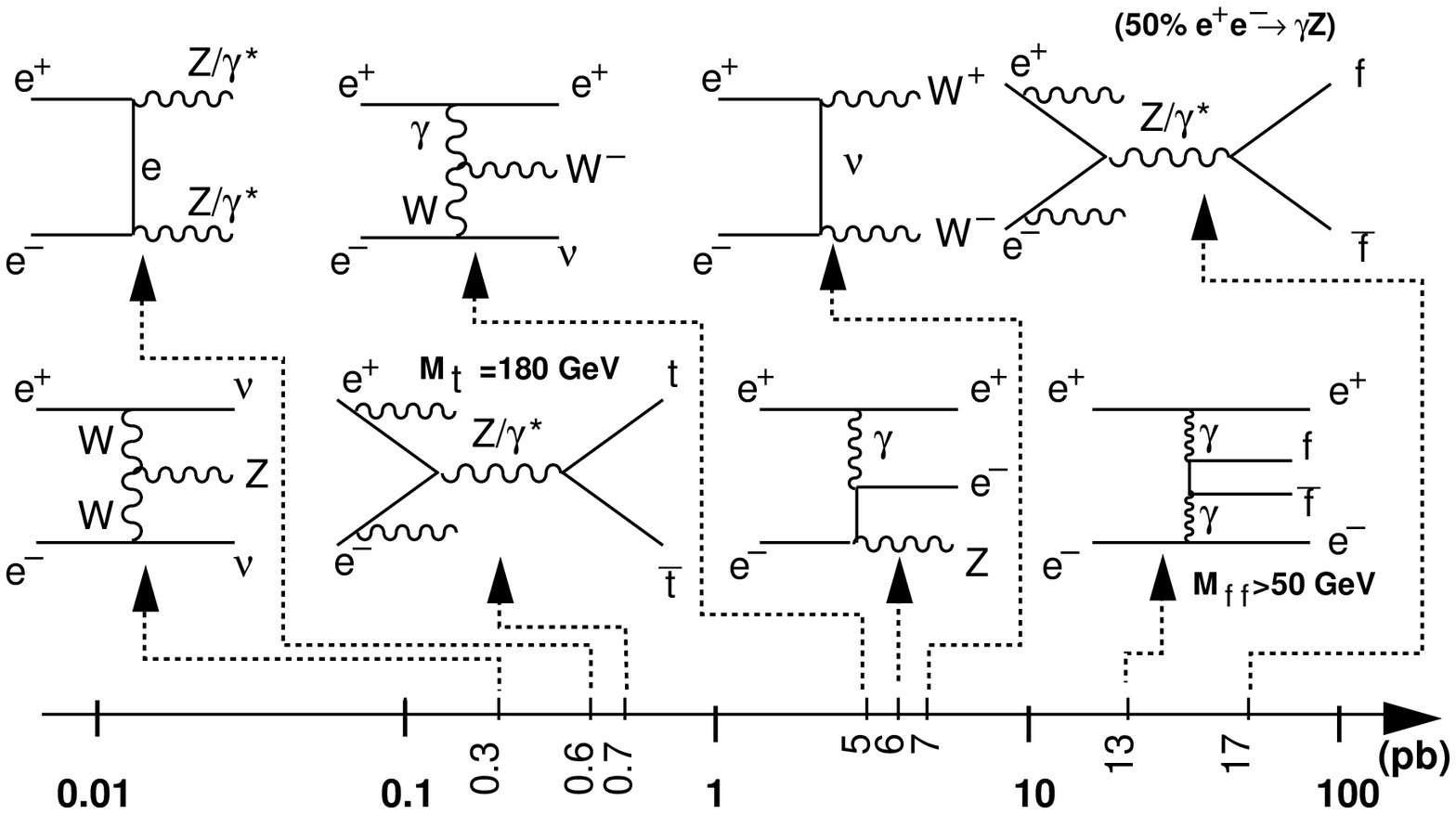, width=14cm}}}
\end{picture}}
\label{fig:bg}
\bce{\large{\bf Fig.~\arabic{figure}}}\ece
\refstepcounter{figure}

\clearpage 

\vspace*{3cm}
\noi 
\begin{minipage}[t]{73mm}   
{\setlength{\unitlength}{1mm}
\begin{picture}(70,100)                        
\put(-2,0){\mbox{\epsfig{figure=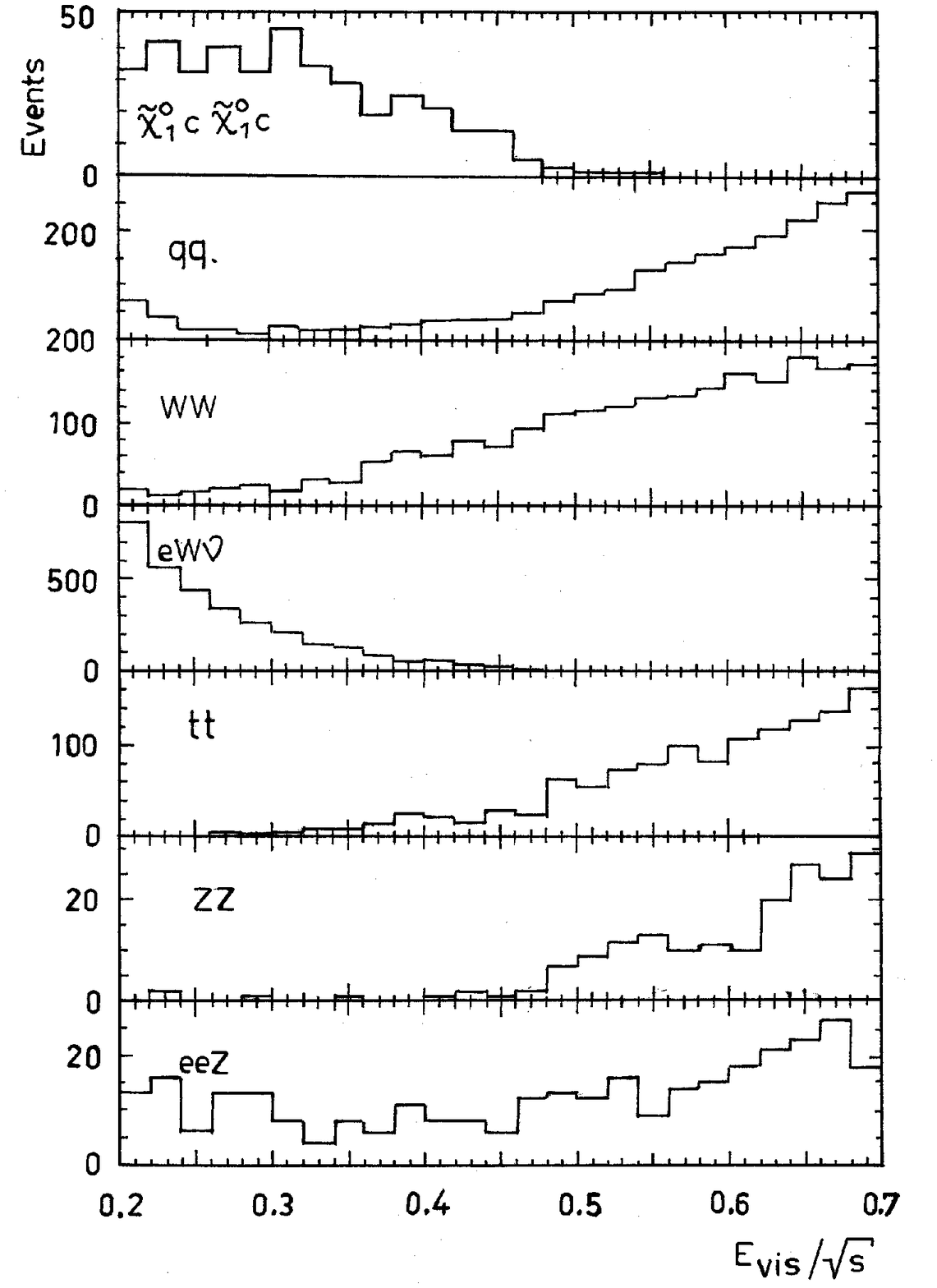,width=7.5cm}}}
\end{picture}}
\bce{\large{\bf Fig.~\arabic{figure}}}\ece \label{fig:evis}
\refstepcounter{figure}
\end{minipage}
  \hspace{3mm}
\begin{minipage}[t]{73mm}   
{\setlength{\unitlength}{1mm}
\begin{picture}(70,100)
\put(-2,0){\mbox{\epsfig{figure=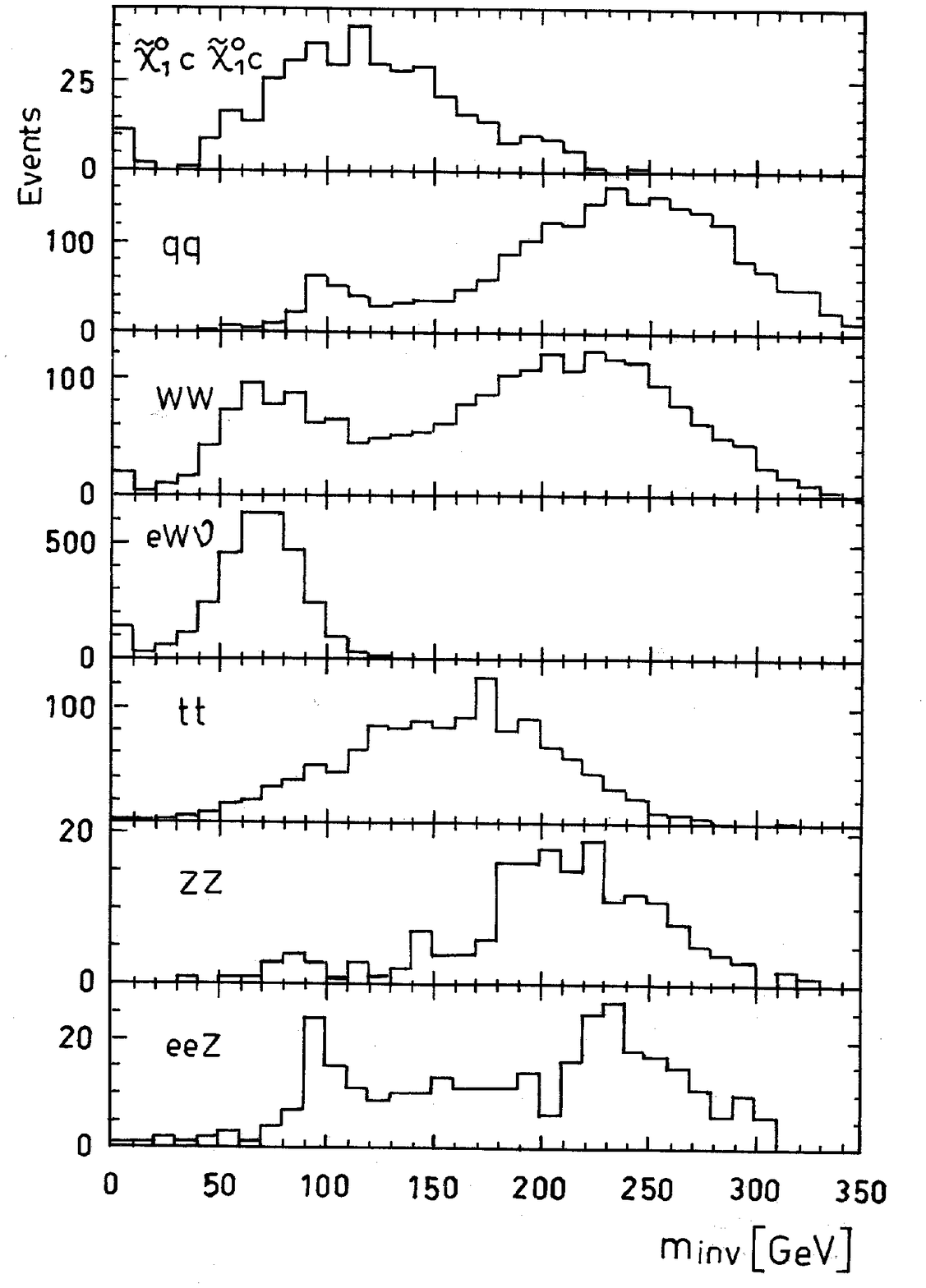,width=7.5cm}}}
\end{picture}}
\bce{\large{\bf Fig.~\arabic{figure}}}\ece \label{fig:minv}
\refstepcounter{figure}
\end{minipage}

\clearpage 

\bce \vspace*{3cm}
\begin{minipage}[t]{73mm}   
{\setlength{\unitlength}{1mm}
\begin{picture}(70,100)
\put(-4,-2){\mbox{\epsfig{figure=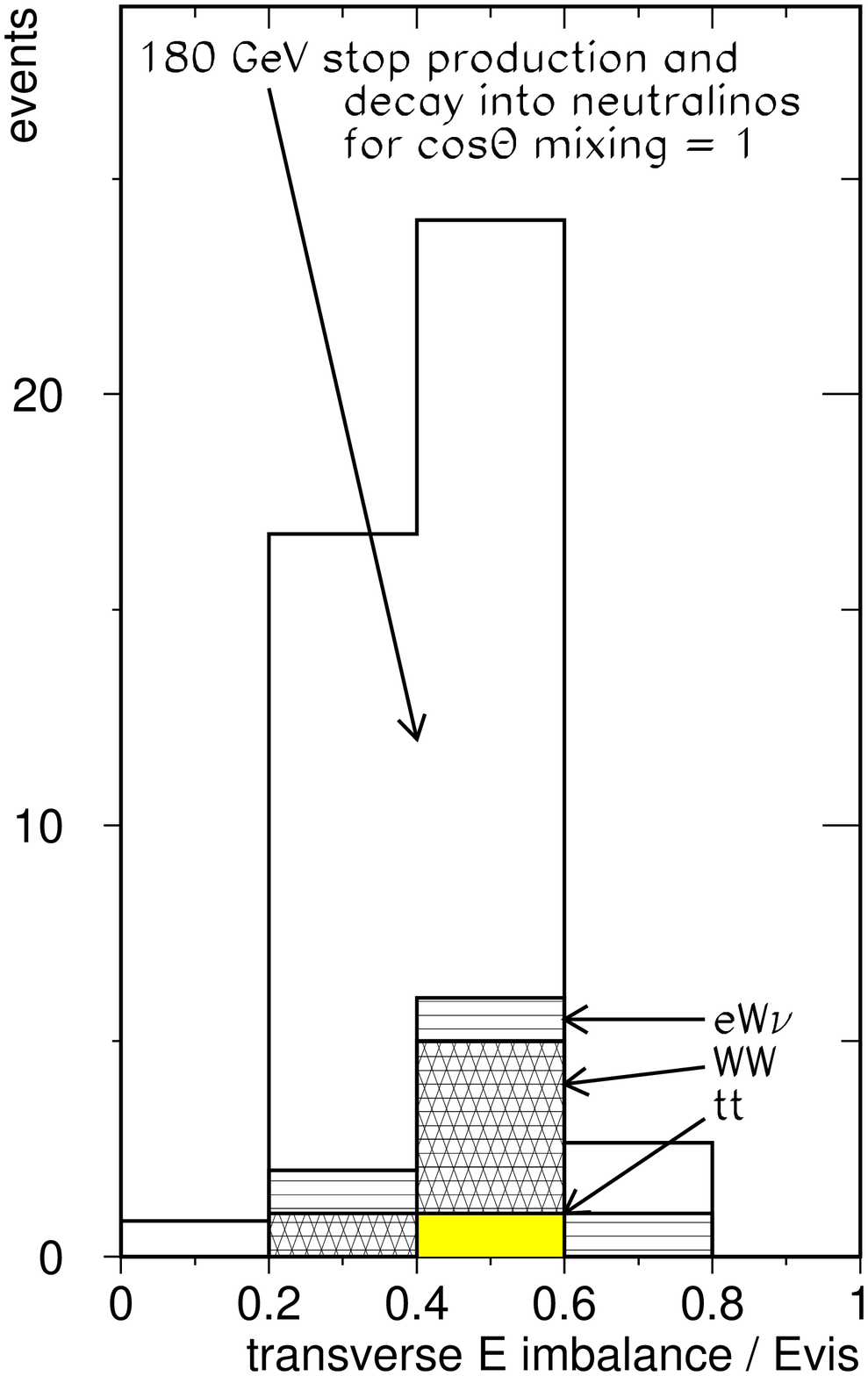,height=10.9cm}}}
\end{picture}}
\bce{\large{\bf Fig.~\arabic{figure}}}\ece  
\refstepcounter{figure}
\end{minipage}
\ece

\clearpage 

\vspace*{3cm}
\noi 
\begin{minipage}[t]{73mm}   
{\setlength{\unitlength}{1mm}
\begin{picture}(70,100)                        
\put(-2,0){\mbox{\epsfig{figure=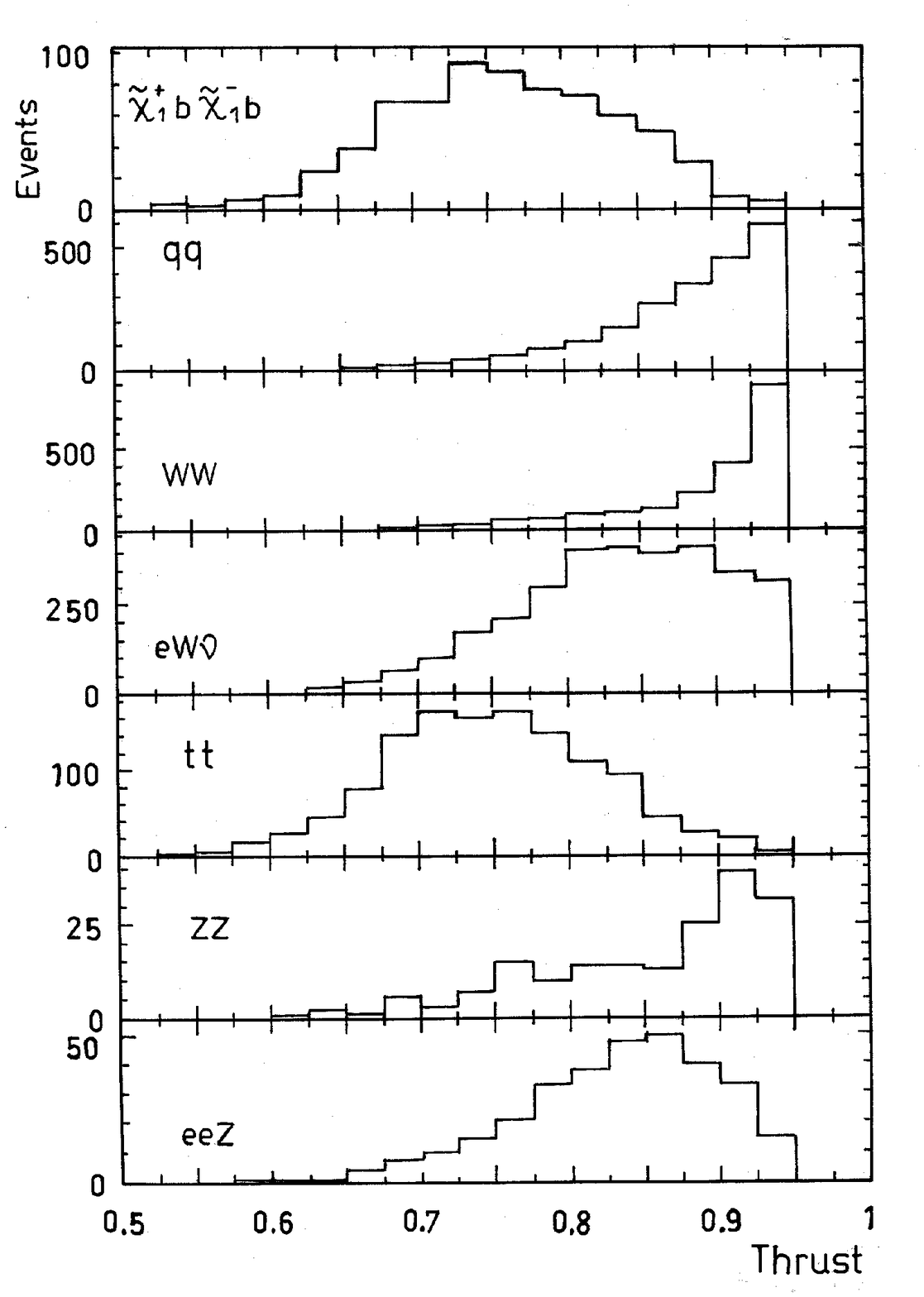,width=7.5cm}}}
\end{picture}}
\bce{\large{\bf Fig.~\arabic{figure}}}\ece \label{fig:thrust}
\refstepcounter{figure}
\end{minipage}
  \hspace{3mm}
\begin{minipage}[t]{73mm}   
{\setlength{\unitlength}{1mm}
\begin{picture}(70,100)
\put(-2,0){\mbox{\epsfig{figure=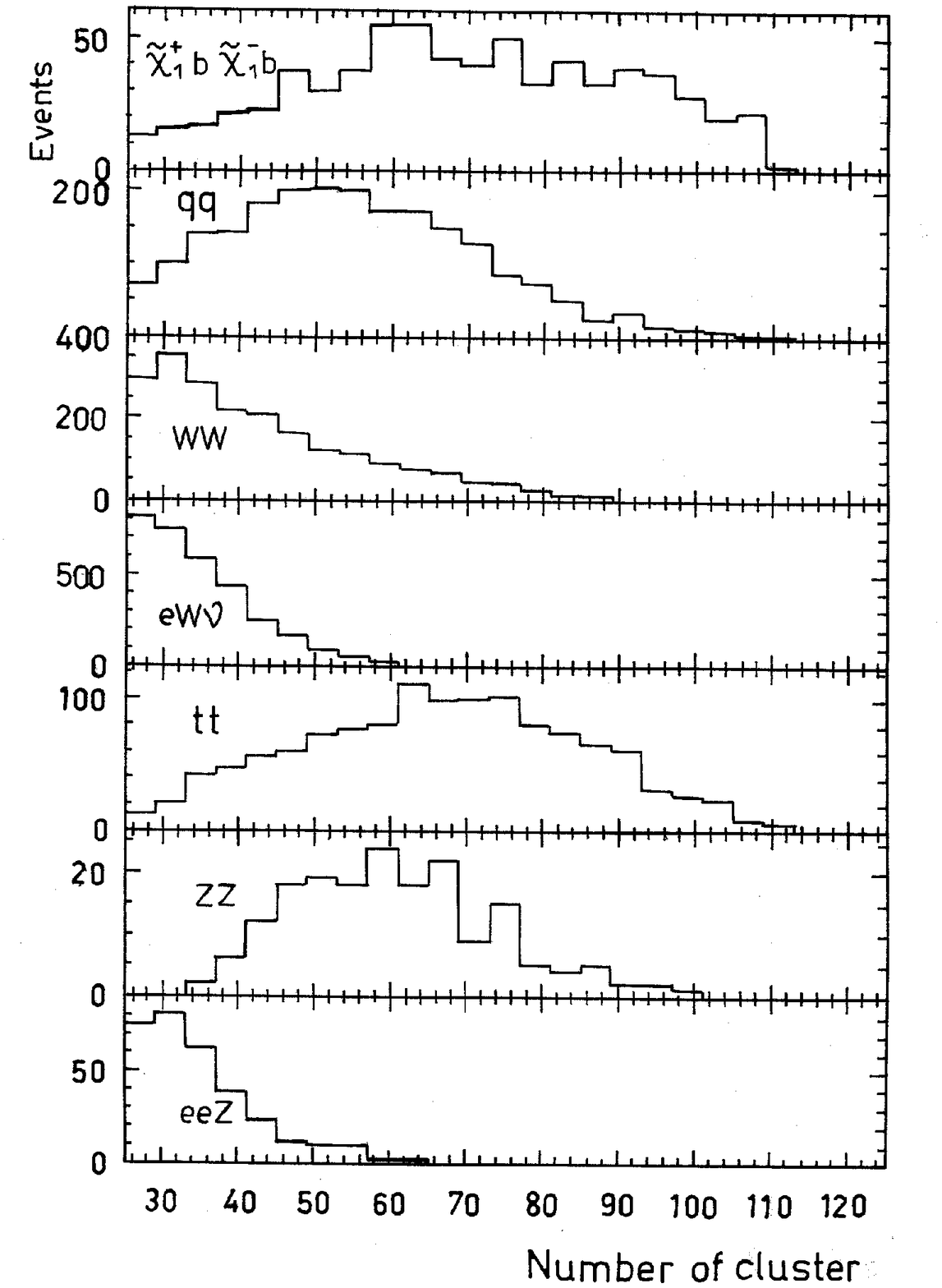,width=7.5cm}}}
\end{picture}}
\bce{\large{\bf Fig.~\arabic{figure}}}\ece \label{fig:cluster}
\refstepcounter{figure}
\end{minipage}
\vspace{15mm}

\clearpage 

\bce \vspace*{3cm}
\begin{minipage}[t]{73mm}   
{\setlength{\unitlength}{1mm}
\begin{picture}(70,100)
\put(-4,-3){\mbox{\epsfig{figure=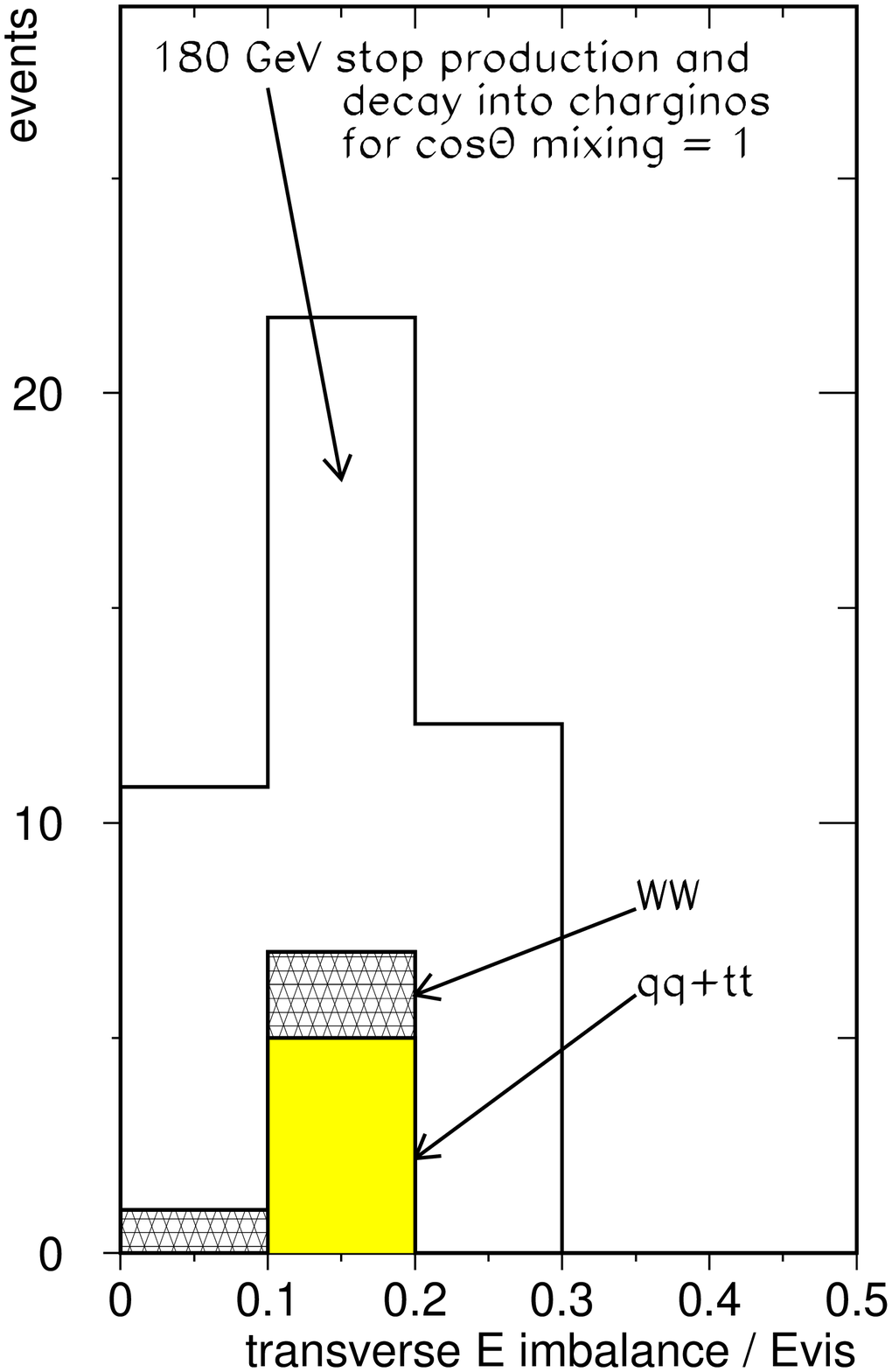,height=10.9cm}}}
\end{picture}}
\bce{\large{\bf Fig.~\arabic{figure}}}\ece  
\refstepcounter{figure}
\end{minipage}
\ece

\clearpage 

\noi
\begin{minipage}[t]{73mm}   
{\setlength{\unitlength}{1mm}
\begin{picture}(70,70)
\put(-2,-3){\mbox{\psfig{figure=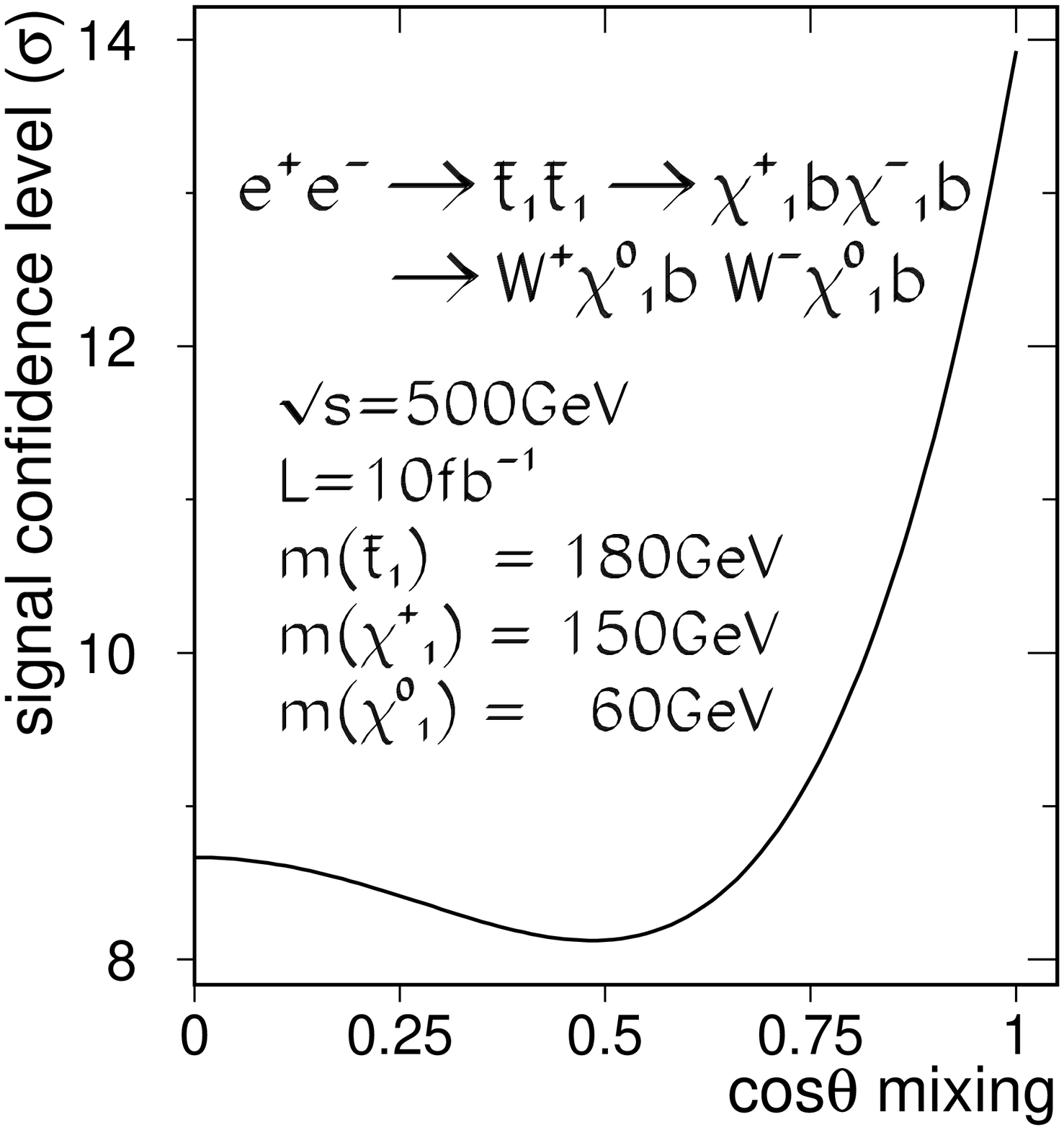,height=8cm}}}
\end{picture} }
\bce{\large{\bf Fig.~\arabic{figure}a}}\ece
\end{minipage} 
  \hspace{3mm}
\begin{minipage}[t]{73mm}    
{\setlength{\unitlength}{1mm}
\begin{picture}(70,70)
\put(-2,-3){\mbox{\psfig{figure=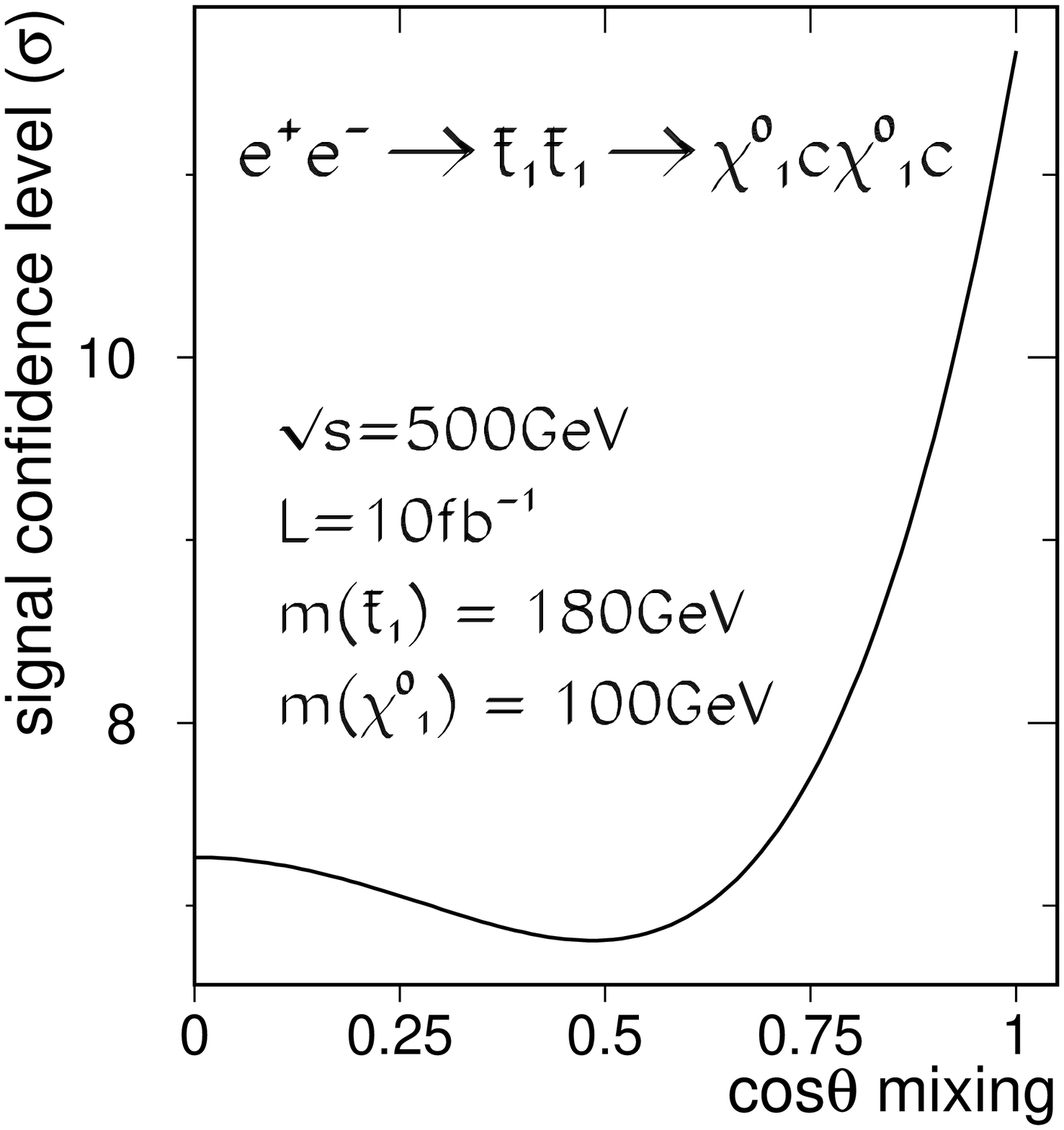,height=8cm}}}
\end{picture} }
\bce{\large{\bf Fig.~\arabic{figure}b}}\ece
\end{minipage} 
\refstepcounter{figure}
\vspace{20mm}

\noi
\begin{minipage}[t]{73mm}   
{\setlength{\unitlength}{1mm}
\begin{picture}(70,70)
\put(0,0){\mbox{\psfig{figure=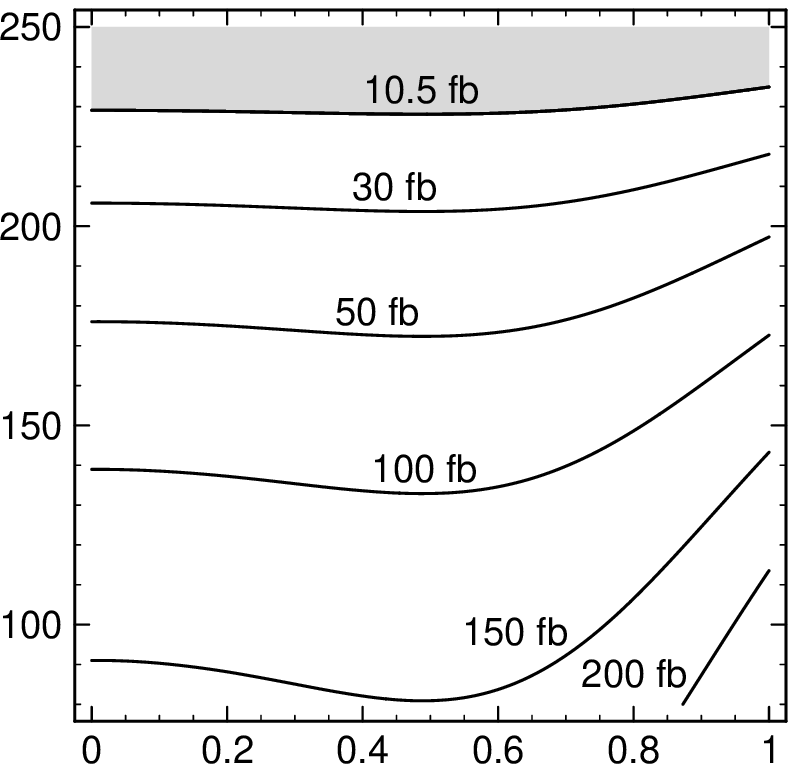,height=7cm}}}
\put(72,0.5){\makebox(0,0)[tr]{$\cos\theta_{\ti t}$}}
\put(-1,71){\makebox(0,0)[bl]{$m_{\ti t_1}$~[GeV] }}
\end{picture} }
\bce{\large{\bf Fig.~\arabic{figure}}}\ece
\end{minipage} 
\refstepcounter{figure}
  \hspace{3mm}
\begin{minipage}[t]{73mm}    
{\setlength{\unitlength}{1mm}
\begin{picture}(70,70)
\put(0,0){\mbox{\psfig{figure=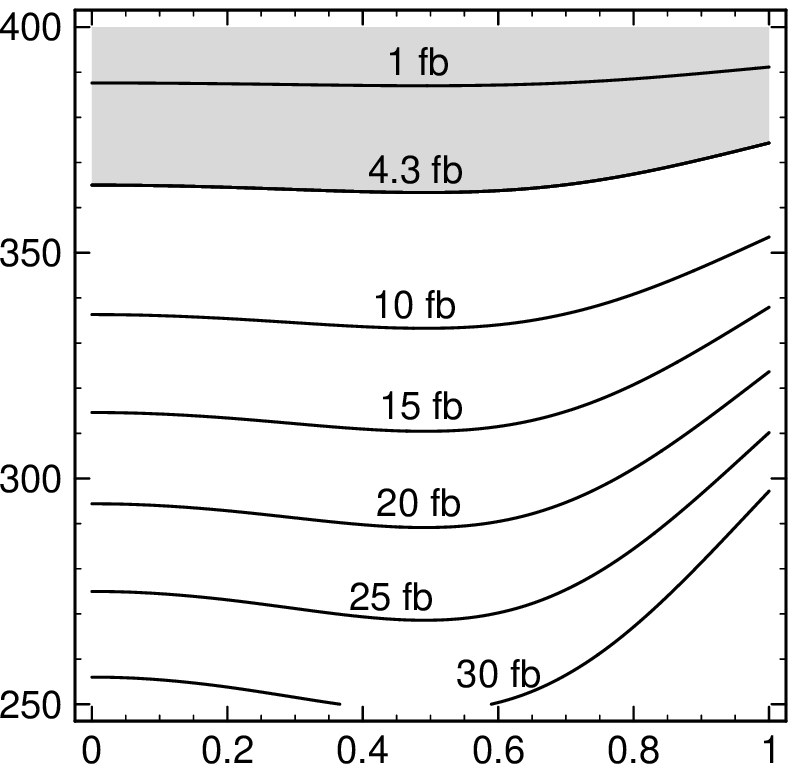,height=7cm}}}
\put(72,0.5){\makebox(0,0)[tr]{$\cos\theta_{\ti t}$}}
\put(-1,71){\makebox(0,0)[bl]{$m_{\ti t_1}$~[GeV] }}
\end{picture} }
\bce{\large{\bf Fig.~\arabic{figure}}}\ece
\end{minipage} 
\refstepcounter{figure}
\vspace{15mm}

\bce

\noindent
{\setlength{\unitlength}{1mm}
\begin{minipage}[t]{8cm}
\begin{picture}(70,70)
\put(0,0){\mbox{\psfig{figure=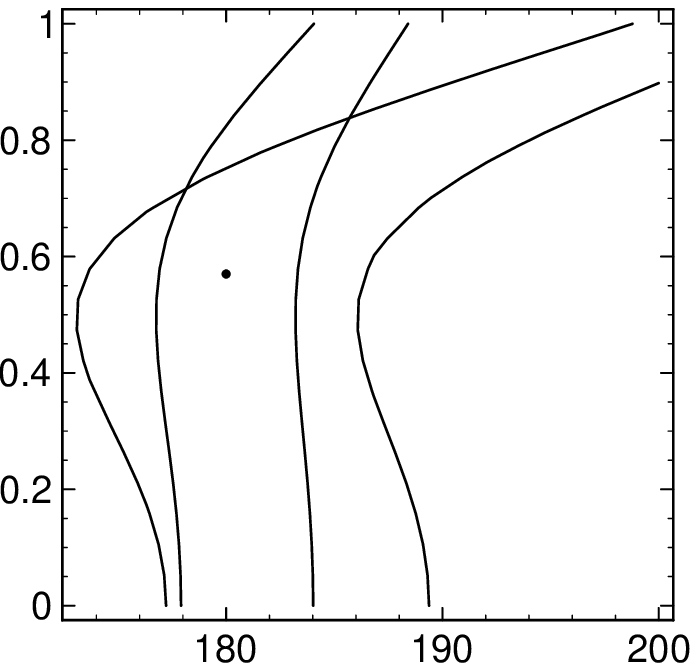,height=7cm}}}
\put(72,0){\makebox(0,0)[tr]{$m_{\st_1}$~[GeV]}}
\put(-1,71){\makebox(0,0)[bl]{$|\cos\theta_{\st}|$}}
\put(37,57){\rotatebox{18}{
            \makebox(0,0)[tl]{{$\sigma_{500} + \Delta \sigma_{500}$}}}}
\put(43,47){\rotatebox{29}{
            \makebox(0,0)[tl]{{$\sigma_{500} - \Delta \sigma_{500}$}}}}
\put(18.5,9){\rotatebox{92}{
    \makebox(0,0)[tl]{$\sigma_{400} + \Delta \sigma_{400}$}}}
\put(31,9){\rotatebox{94}{
    \makebox(0,0)[bl]{$\sigma_{400} - \Delta \sigma_{400}$}}}
\end{picture}
\bce{\large{\bf Fig.~\arabic{figure}}}\ece
\end{minipage} }
\refstepcounter{figure}
  
  \vspace{15mm}
  
{\setlength{\unitlength}{1mm}
\begin{minipage}[t]{8cm}
\begin{picture}(70,70)
\put(0,0){\mbox{\psfig{figure=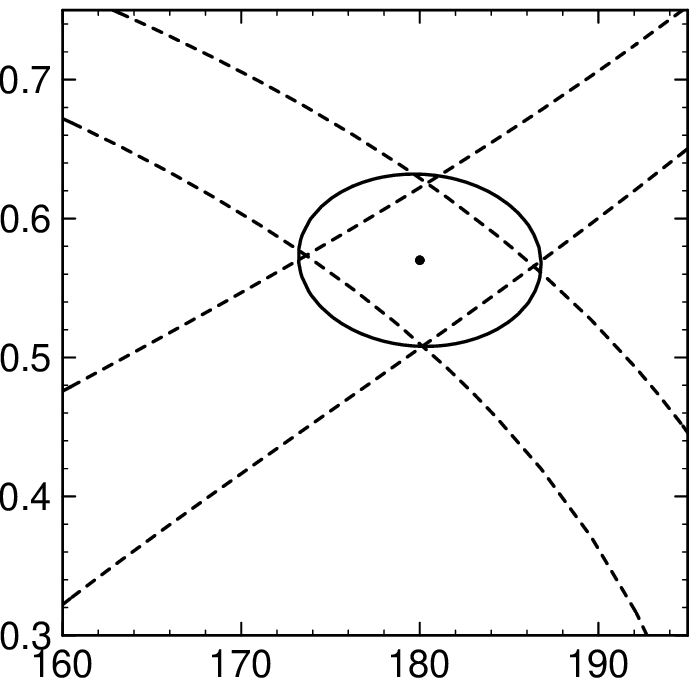,height=7cm}}}
\put(72,0){\makebox(0,0)[tr]{$m_{\st_1}$~[GeV]}}
\put(-1,71){\makebox(0,0)[bl]{$|\cos\theta_{\st}|$}}
\put(49.5,51){\makebox(0,0)[bl]{68\%~CL}}
\put(9,32.5){\rotatebox{31}{
            \makebox(0,0)[bl]{{$\sigma_L + \Delta \sigma_L$}}}}
\put(17,17.5){\rotatebox{37}{
            \makebox(0,0)[bl]{{$\sigma_L - \Delta \sigma_L$}}}}
\put(11,57){\rotatebox{-30}{
    \makebox(0,0)[bl]{$\sigma_R + \Delta \sigma_R$}}}
\put(23,64.5){\rotatebox{-29}{
    \makebox(0,0)[bl]{$\sigma_R - \Delta \sigma_R$}}}
\end{picture}
\bce{\large{\bf Fig.~\arabic{figure}}}\ece
\end{minipage} 
}
\refstepcounter{figure}

\ece

\end{document}